\documentclass[english,journal=jcpafh]{achemso}
\usepackage[T1]{fontenc}
\usepackage[utf8]{inputenc}
\usepackage{color}
\usepackage{amsbsy}
\usepackage{amstext}
\usepackage{amssymb}
\usepackage{graphicx}
\usepackage{float}

\makeatletter

\title{Benchmarking Gaussian Basis Sets in Quantum-Chemical Calculations
of Photoabsorption Spectra of Light Atomic Clusters}

\author{Vikram Mahamiya}

\email{mahamiyavikram@gmail.com}

\affiliation{Department of Physics, Indian Institute of Technology Bombay, Powai,
Mumbai 400076, India}

\author{Pritam Bhattacharyya}

\email{pritambhattacharyya01@gmail.com}

\affiliation{Department of Physics, Indian Institute of Technology Bombay, Powai,
Mumbai 400076, India}

\altaffiliation{Present Address: Institute for Theoretical Solid State Physics, Leibniz
IFW Dresden, Helmholtzstr. 20, 01069 Dresden, Germany}

\author{Alok Shukla}

\email{shukla@iitb.ac.in}

\affiliation{Department of Physics, Indian Institute of Technology Bombay, Powai,
Mumbai 400076, India}

\providecommand{\tabularnewline}{\\}

\makeatother

\usepackage{babel}
\begin{document}
\begin{abstract}
The choice of Gaussian basis functions for computing the ground-state
properties of molecules, and clusters, employing wave-function-based
electron-correlated approaches, is a well-studied subject. However,
the same cannot be said when it comes to the excited-state properties
of such systems, in general, and optical properties, in particular.
The aim of the present study is to understand how the choice of basis
functions affects the calculations of linear optical absorption in
clusters, qualitatively, and quantitatively. For this purpose, we
have calculated linear optical absorption spectra of several small
charged and neutral clusters, namely, Li$_{2}$, Li$_{3}$, Li$_{4}$,
B$_{2}^{+}$, B$_{3}^{+}$, Be$_{2}^{+}$, and Be$_{3}^{+}$, using
a variety of Gaussian basis sets. The calculations were performed
within the frozen-core approximation, and a rigorous account of electron
correlation effects in the valence sector was taken by employing various
levels of configuration interaction (CI) approach both for the ground
and excited states. Our results on the peak locations in the absorption
spectra of Li$_{3}$ and Li$_{4}$ are in very good agreement with
the experiments. Our general recommendation is that for excited-state
calculations, it is very important to utilize those basis sets which
contain augmented functions. Relatively smaller aug-cc-pVDZ basis
sets also yield high-quality results for photoabsorption spectra,
and are recommended for such calculations if the computational resources
are limited. 
\end{abstract}

\section{Introduction}

\label{sec:introd}

Gaussian basis functions (GBFs) were initially proposed by Boys for
use in computational atomic and molecular quantum mechanics,\cite{boys-gaussian}
and over the years have become the preferred basis functions in quantum
chemistry.\cite{boys-gaussian} The reason behind the popularity of
GBFs is the so-called Gaussian product theorem\cite{boys-gaussian,mcmurchie-davidson}
which allows for analytical results for the expressions of multi-center
integrals involving various physical quantities. Nevertheless, one
has to be always careful about various convergence related issues
when using GBFs, because, unlike Slater basis functions, they do not
exhibit correct asymptotic behavior far away from the nuclei. This
generally leads to the requirement that a large number of GBFs should
be used to achieve convergence, leading to huge memory and CPU-time
requirements because the required number of integrals scale as $\approx N^{4}$,
where $N$ is the total number of basis functions. Keeping this in
mind, several groups have studied the convergence properties of GBFs
over the years, and have come up with schemes to balance accuracy
with the computational effort (see, e.g., Refs. \cite{basis-set-review-davidson-feller,huzinaga-basis-set-review}
for comprehensive reviews). Huzinaga was one of the earliest researchers
to optimize GBFs for Hartree-Fock (HF) calculations on atoms.\cite{huzinaga1}
Ruedenberg and coworkers devised the so-called even-tempered basis
set,\cite{even-tempered1,even-tempered2} while Huzinaga and coworkers
developed well-tempered basis functions.\cite{huzinaga-wt} Huzinaga
and coworkers further developed several contracted basis sets,\cite{tatewaki-huzinaga-1979,tatewaki-huzinaga-1980}
discussed in detail in Ref.\cite{huzinaga-basis-set-review} Pople
and coworkers developed a large number of basis sets\cite{huzinaga-basis-set-review,basis-set-review-davidson-feller}
which enjoy continued popularity even in present times. One of the
most popular minimal basis sets introduced by Pople and coworkers
is STO-3G contracted basis set,\cite{pople-sto-ng} whose purpose
was to emulate Slater-type orbitals, using GTFs. Split-valence basis
sets are among the most popular extended basis sets introduced by
Pople et al.,\cite{pople-431g,pople-321g,pople-621g,pople-6311g-6311g*}
in which for inner shells contracted minimal basis functions are used,
but for the valence shells a split set of basis functions is employed,
which consist of both contracted and primitive GTFs. Depending upon
the contraction schemes, these basis sets were given names such as
3-21G,\cite{pople-321g} 4-31G,\cite{pople-431g} 6-21G,\cite{pople-621g}
and 6-311G\cite{pople-6311g-6311g*}, etc. Pople and coworkers also
proposed further enlarged basis sets containing polarization and diffuse
functions of higher angular momenta, which have since become popular
choices in quantum chemistry.\cite{pople1973-631g*,pople-sto3g*,pople-6311g-6311g*,pople-631g*-second-row,pople-321g*,pople-6311g*-larger}
Dunning and coworkers introduced a series of extended basis sets,
called ``correlation-consistent'' (CC) basis sets, which are of
varying sizes, containing both polarization and diffuse functions.\cite{DUNNING_1,DUNNING_2,DUNNING_3}
The basic idea behind these CC basis sets is that they recover a significant
amount of electron correlation energy in post-Hartree-Fock treatments
of corresponding atoms. In addition to the basis sets mentioned here,
numerous other sets of basis functions have been developed over the
years, for which we refer the reader to review articles by Davidson
and Feller,\cite{basis-set-review-davidson-feller} and Huzinaga.\cite{huzinaga-basis-set-review} 

Even though so many basis sets have been developed by numerous groups, in most of the reports
the criteria for their selection appears to be driven by a good description
of the ground state energies of the atoms involved either at the Hartree-Fock
level, or in electron-correlated calculations.\cite{huzinaga-basis-set-review,basis-set-review-davidson-feller}
Previously, Balakina et al.\cite{BALAKINA}  have explored the basis set dependence of the linear and non-linear optical properties of conjugated organic molecule p-nitroaniline. They reported that the [4s3p2d/3s] basis set also provides similar results as aug-cc-pVDZ basis set for the calculations of (hyper)polarizability. Parsons et al.\cite{Parsons} have explored the basis set dependence of optical rotation calculations of various types of gauges. They found that the origin-invariant length gauge (LG-OI) gauge with aug-cc-pVTZ basis set provides a balance of cost and accuracy for DFT method. Reis and Papadopoulos\cite{reis} reported that the inclusion of f-functions in the Dunning’s basis sets does not have a large effect on the electric properties of B4 cluster. Lauderdale and Coolidge\cite{Lauderdale} have explored the effect of basis sets on the non-linear optical properties (hyperpolarizabilities) of linear diacetylenes using time-dependent Hartree-Fock theory. They found that the inclusion of a diffuse ‘d’ function to a standard double-zeta plus polarization basis can significantly improve the frequency-dependent hyperpolarizability. Jabłonski and Palusiak\cite{Jablonski} have explored the influence of basis sets in Hartree-Fock (HF) and DFT/B3LYP calculations for the values atoms in molecules (AIM) parameters. They found that smaller Dunning’s basis sets, including cc-pVDZ and aug-cc-pVDZ provide poor results as compared to medium-sized Pople-type basis sets. We are not aware of a systematic study in which the basis sets have
been examined from the perspective of their performance in excited
state calculations. Furthermore, we have also not come across a study
which examines the basis sets from the point of view their ability
to compute optical properties of atoms and molecules, which involves
calculations of transition dipole moments, in addition to excited
state energies, and wave functions. In order to fill this void, we
decided to undertake a systematic investigation of the influence of
basis sets on the qualitative and quantitative description of optical
absorption spectra of atomic clusters. In this paper, we have performed
calculations of linear optical absorption spectra of several small
neutral and cationic clusters, e.g., Li$_{2}$, Li$_{3}$, Li$_{4}$,
B$_{2}^{+}$, B$_{3}^{+}$ , Be$_{2}^{+}$, and Be$_{3}^{+}$ , using
the configuration-interaction (CI) approach. For this purpose, a number
of basis sets, namely, 6-311++G(2d,2p), 6-311++G(3df,3pd), cc-pVDZ,
cc-pVTZ, aug-cc-pVDZ, and aug-cc-pVTZ, were employed, and their influence
on the convergence of excited state energies, wave functions, and
transition dipole moments has been systematically examined. In this
study, the reason behind our choice of smaller sized atomic clusters and their ions, as
against larger ones, is that it is possible to perform highly accurate
CI calculations on smaller systems so that the difference between
results obtained with different basis sets will be due the nature
of basis sets, and not due to the CI approach employed. Based upon
our calculations, the main conclusion is that it is very important
to include diffuse basis functions in the basis set in order to obtain
a good description of the photoabsorption spectra.

\section{Theoretical Approach and Computational Details}

\subsection{General Methodology\label{sec:theory}}

All the calculations were performed using the first-principles wave-function-based
electron-correlated approaches, using the standard Hamiltonian within
the Born-Oppenheimer approximation. The molecular orbitals are expressed
in terms of the linear combination of Cartesian-Gaussian type basis
functions, also called atomic orbitals (AOs). Although, for such calculations,
a number of program packages are available, we employed GAUSSIAN16\cite{g16}
and MELD\cite{meld} for our calculations. The geometries of all the
clusters considered in this work were optimized using GAUSSIAN16 package\cite{g16}
at the coupled-clusters singles-double (CCSD) level of theory, employing
a large augmented correlation-consistent polarized valence triple-zeta
(aug-cc-pVTZ) basis set.

We perform excited-states calculations for various clusters employing
their ground-state optimized geometries, using the configuration-interaction
(CI) methodology at various levels of approximation, as implemented
in the program package MELD\cite{meld}. The CI calculations yield
the vertical excitation energies, the ground and excited state wave
functions, and the transition dipole matrix elements connecting the
ground and the excited states, which, in turn, are used to compute
the optical absorption spectra of various clusters. The level of CI
employed in the calculations depends on the size of cluster, the number
of valence electrons in cluster, and the number of active orbitals.
The linear optical absorption spectra of Li$_{2}$, Li$_{3}$, Li$_{4}$,
Be$_{2}^{+}$, B$_{2}^{+}$ clusters were computed at the full CI
(FCI) level, while for Be$_{3}^{+}$ and B$_{3}^{+}$ calculations
were performed at the quadruple CI (QCI), and the multi-reference
singles-doubles CI (MRSDCI) levels, respectively. 

We start the calculations on a given cluster by first performing restricted
Hartree-Fock (RHF) calculations on it, and obtain the molecular orbitals
(MOs), expressed as linear combinations of the chosen AOs. In order
to perform CI calculations, the one- and two-electron Hamiltonian
matrix elements are transformed from the AO representation to the
MO representation. For the FCI calculations, all possible configurations
obtained by placing all the valence electrons of the cluster in the
given set of MOs, in all possible ways, consistent with the Pauli
exclusion principle. In the QCI approach, we first choose a reference
configuration, and then generate configurations which are singly-,
doubly-, triply-, and quadruply-excited with respect to it. For the
ground-state calculations, the reference configuration is normally
taken to be the RHF configuration, while for the excited-state calculations
one chooses an excited configuration which is closest to the excited
state one is trying to calculate. However, both the FCI and the QCI
approaches can lead to a very large number of configurations if the
number of electrons and the MO basis is large, thus, making the calculations
intractable. Therefore, for the larger clusters, we employed the multi-reference
singles-doubles configuration-interaction (MRSDCI) approach, as implemented
in the MELD package. In this approach, the singly- and doubly-excited
configurations are generated from a list of configurations called
the reference configurations, chosen by the user. We performed the
MRSDCI calculations in an incremental manner, by starting out with
a small set of reference configurations that are close to the states
(ground or excited) we are targeting. Then we analyze the optical
absorption spectra of the cluster calculated from that MRSDCI calculation,
and identify a new set of configurations which need to be included
in the list of reference configurations based upon their contributions
in the wave functions of the targeted states. The procedure is iterated
until the calculated optical absorption spectrum converges to within
a user-defined threshold. In all the CI calculations, the configurations
are actually configuration-state functions (CSFs) which are eigenstates
of the point-group symmetry operators, and the total spin operators
$S^{2}$ and $S_{z}$.\cite{C4CP02232G,doi:10.1021/acs.jpcc.7b08695,doi:10.1021/jp408535u,doi:10.1063/1.4867363,doi:10.1063/1.4897955,doi:10.1142/S1793984411000529,PhysRevB.65.125204,PhysRevB.69.165218,PhysRevB.75.155208,PhysRevB.92.205404,Priya2017,Shinde2017}

The linear optical absorption spectrum of a given cluster is calculated
under the electric-dipole approximation, using the formula 
\begin{equation}
\sigma(\omega)=4\pi\alpha\sum_{i}\frac{\omega_{io}|\langle i|\boldsymbol{\hat{e}.r}|0\rangle|^{2}\gamma^{2}}{(\omega_{i0}-\omega)^{2}+\gamma^{2}}\label{eq:optics}
\end{equation}

Above: (i) $\sigma(\omega)$ represents the optical absorption cross
section, (ii) $\omega$ is the frequency of incident light, (iii)$\hat{\boldsymbol{e}}$
denotes the polarization direction of the incident light, (iv) $\boldsymbol{r}$
is the position operator, (v) $\alpha$ is the fine structure constant,
(vi) $\hbar\omega_{i0}$ is the energy difference between ground state
(0) and the $i^{th}$ excited state $(i)$, and (vii) $\gamma$ is
the uniform line width associated with each excited state. The line
width $\gamma$ is taken to be 0.1 eV in all our calculations. The
sum over index $i$ denotes the sum over all possible excited states.
We have restricted this sum in our calculations up to the states corresponding
to excitation energies of 10 eV, or less. Additionally, the oscillator
strength $f_{n}$ corresponding to an optical transition from the
ground state to the $n$-th excited state is computed using the standard
formula 
\begin{equation}
f_{n}=\frac{2m_{e}}{3\hbar^{2}}\Delta E_{n}\sum_{j=x,y,z}\sum_{\alpha}\arrowvert\langle n\alpha\arrowvert O_{j}\arrowvert0\rangle\arrowvert^{2}\label{eq:osc-strength}
\end{equation}
 above $m_{e}$ is the electron mass, $|0\rangle$ and $|n\alpha\rangle$
are, respectively the CI wave functions of the ground state and the
excited state in question, with $\alpha$ being a degeneracy label,
$O_{j}$ denotes $j$-th Cartesian component of the electric-dipole
operator, while $\Delta E_{n}=E_{n}-E_{0}$ is the excitation energy
of the excited state. 

\subsection{Computational Parameters}

In this section, we will discuss the convergence of the results with
respect two parameters, related to the basis-set-size: (a) number
of active orbitals in the CI calculations, and (b) number of CSFs
included in the calculations.

\subsubsection{Active molecular orbitals}

It is well-known that the computational cost at configuration interaction
(CI) level of theory increases as $N_{act}^{6}$, where $N_{act}$
is the total number of active molecular orbitals used in the CI calculations.
Therefore, the time needed to perform a CI calculation will proliferate
rapidly with the increasing values of $N_{act}$. We have adopted
two approaches to reduce the size of the active MO set: (a) we adopt
the frozen-core approximation to eliminate the core orbitals of each
atom of the cluster, and (b) for certain cases involving large CI
matrices, we delete all those virtual (unoccupied) orbitals from our
calculations whose single-particle energies are larger than 1 Hartree.
The frozen-core approximation is a standard approach which also has
the added advantage of considerably reducing the number of active
electrons ($n_{elec}$) in the calculation. The ``1 Hartree cutoff''
also doesn't reduce the accuracy of the calculations because we are
interested in low-lying optical excitations below 10 eV, while our
cutoff eliminates only those orbitals from the calculations whose
energy is larger than 27.21 eV. Both these approximations have been
investigated rigorously in our group in earlier calculations.\cite{doi:10.1142/S1793984411000529,Shinde2017,Priya2017,Bhattacharyya_2019}.

To be specific, in the present set of calculations, we have considered
all the virtual orbitals for Li$_{2}$ and Be$_{2}^{+}$ clusters,
while for Li$_{3}$, Li$_{4}$, Be$_{3}^{+}$, B$_{2}^{+}$ and B$_{3}^{+}$
clusters we have imposed the 1 Hartree cutoff. 

\subsubsection{Size of CI expansion}

\begin{figure}[H]
	\includegraphics[scale=0.5]{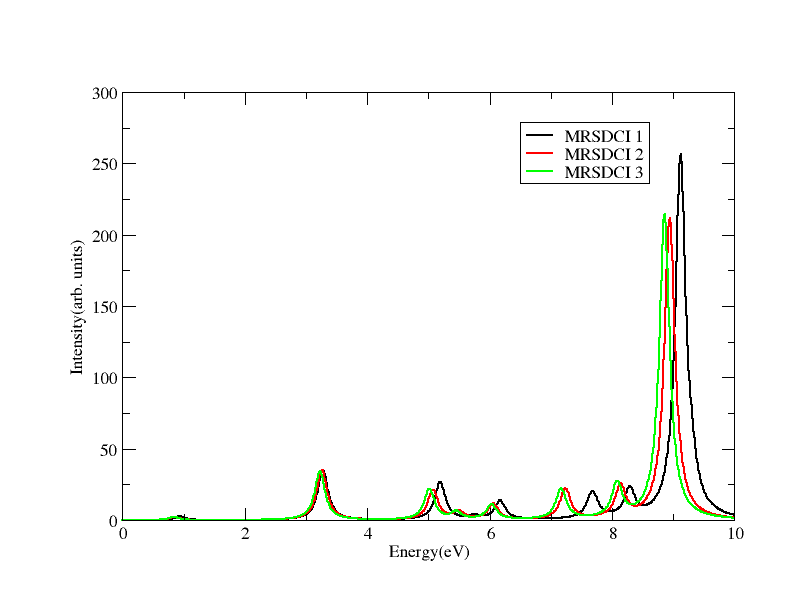}\caption{Convergence of the optical absorption spectrum of the B$_{3}^{+}$
		computed using the MRSDCI method, with the increasing numbers of reference
		configurations ($N_{ref}$). For calculations labeled MRSDCI1, MRSDCI2,
		and MRSDCI3, values of $N_{ref}$ were 58, 101, and 144, respectively.\label{fig:mrsdci-convergence}}
\end{figure}

Another important parameter that controls the quality of calculations
is the total number of CSFs, $N_{total}$, included in the CI expansion
of the many-particle wave functions of the clusters concerned, both
for their ground and the excited states. As mentioned earlier, for
a given set of active electrons and MOs, the best possible CI expansion
corresponds to the FCI expansion, which becomes intractable for systems
with large values of $n_{elec}$ and $N_{act}$. However, whenever
FCI is not possible, we employ one of the restricted CI approaches
such as the QCI or the MRSDCI methods. Of the two, it is crucial to
examine the convergence of the MRSDCI approach which is based upon
singles and doubles excitations from a number of reference configurations
($N_{ref}$) leading to the final CI expansion with $N_{total}$ CSFs.
We examined the convergence of the optical absorption spectrum for
the B$_{3}^{+}$ cluster calculated using the MRSDCI method, with
respect to $N_{ref}$, and $N_{total}$, as presented in Fig. \ref{fig:mrsdci-convergence}.

In the figure, we plot the absorption spectra of B$_{3}^{+}$ obtained
from three MRSDCI calculations of increasing sizes labeled as MRSDCI1,
MRSDCI2, and MRSDCI3. In these calculations, the values of parameters
$N_{ref}$ and $N_{total}$ were $N_{ref}$= 58, $N_{total}$= 4007873,
$N_{ref}$= 101, $N_{total}$= 5781436, and $N_{ref}$= 144, $N_{total}$=
8422193, respectively. From Fig. \ref{fig:mrsdci-convergence}., it
is obvious that the spectra obtained using MRSDCI2 and MRSDCI3 calculations
are very close to each other, signaling convergence with respect to
the size of the MRSDCI expansion. 

\section{Results and Discussion \label{sec:Results-and-discussion}}

Before discussing the results of our calculations of the optical absorption
spectra of various clusters, we first summarize their ground state
geometries in Table \ref{tab:Point-group,-symmetry}, optimized at
the CCSD level of theory, employing GAUSSIAN16 suite of programs\cite{g16},
and large aug-cc-pVTZ basis sets. 

\begin{table}
	\caption{The nature of the structure, along with the point group symmetry utilized,
		during the coupled-cluster singles-doubles (CCSD) geometry optimization
		calculations are presented below. Additionally, for each cluster,
		the symmetry of the ground-state wave function, total Hartree-Fock
		(HF) energy in Hartree (Ha), total CCSD energy (Ha), and the correlation
		energy (eV) are also presented. During the calculations, aug-cc-pVTZ
		basis set was employed for each cluster.\label{tab:Point-group,-symmetry}}
	
	\centering{}%
	\begin{tabular}{|c|c|c|c|c|c|c|}
		\hline 
		{\scriptsize{}Cluster} & {\scriptsize{}Structure} & {\scriptsize{}Point group} & {\scriptsize{}Symmetry of the} & {\scriptsize{}HF energy} & {\scriptsize{}CCSD energy} & {\scriptsize{}Correlation energy}\tabularnewline
		&  &  & {\scriptsize{}GS wave function} & {\scriptsize{}(Ha)} & {\scriptsize{}(Ha)} & {\scriptsize{}(eV)}\tabularnewline
		\hline 
		\hline 
		{\scriptsize{}Li$_{2}$} & {\scriptsize{}Linear} & {\scriptsize{}D$_{2h}$} & {\scriptsize{}$^{1}$A$_{g}$} & {\scriptsize{}-14.8715509} & {\scriptsize{}-14.9033549} & {\scriptsize{}0.87}\tabularnewline
		\hline 
		{\scriptsize{}Li$_{3}$} & {\scriptsize{}Linear} & {\scriptsize{}D$_{2h}$} & {\scriptsize{}$^{2}$A$_{g}$} & {\scriptsize{}-22.3088776} & {\scriptsize{}-22.3454346} & {\scriptsize{}0.99}\tabularnewline
		\hline 
		{\scriptsize{}Li$_{3}$} & {\scriptsize{}Isosceles triangle} & {\scriptsize{}C$_{2v}$} & {\scriptsize{}$^{2}$A$_{1}$} & {\scriptsize{}-22.3170594} & {\scriptsize{}-22.3557287} & {\scriptsize{}1.05}\tabularnewline
		\hline 
		{\scriptsize{}Li$_{4}$} & {\scriptsize{}Rhombus} & {\scriptsize{}D$_{2h}$} & {\scriptsize{}$^{1}$A$_{g}$} & {\scriptsize{}-29.7619144} & {\scriptsize{}-29.8354840} & {\scriptsize{}2.00}\tabularnewline
		\hline 
		{\scriptsize{}Be$_{2}^{+}$} & {\scriptsize{}Linear} & {\scriptsize{}D$_{2h}$} & {\scriptsize{}$^{2}$A$_{g}$} & {\scriptsize{}-28.9205835} & {\scriptsize{}-28.9672583} & {\scriptsize{}1.27}\tabularnewline
		\hline 
		{\scriptsize{}Be$_{3}^{+}$} & {\scriptsize{}Linear} & {\scriptsize{}D$_{2h}$} & {\scriptsize{}$^{2}$A$_{g}$} & {\scriptsize{}-43.5410215} & {\scriptsize{}-43.6332325} & {\scriptsize{}2.51}\tabularnewline
		\hline 
		{\scriptsize{}B$_{2}^{+}$} & {\scriptsize{}Linear} & {\scriptsize{}D$_{2h}$} & {\scriptsize{}$^{2}$A$_{g}$} & {\scriptsize{}-48.8344277} & {\scriptsize{}-48.9626672} & {\scriptsize{}3.49}\tabularnewline
		\hline 
		{\scriptsize{}B$_{3}^{+}$} & {\scriptsize{}Equilateral triangle} & {\scriptsize{}D$_{3h}$} & {\scriptsize{}$^{1}$A$_{1}^{'}$} & {\scriptsize{}-73.4445744} & {\scriptsize{}-73.7127527} & {\scriptsize{}7.27}\tabularnewline
		\hline 
	\end{tabular}
\end{table}

For each cluster, the table lists
the nature of its ground-state structure, point group employed in
the calculations, symmetry of the ground state wave function, total
energy of the ground state, and the correlation energy. 
In Table \ref{tab:ci-calc-details},
the details related to our CI calculations performed for computing
the optical absorption spectra of various clusters are provided. For
various clusters, the table lists: (a) the type of CI calculation,
(b) the point-group symmetry employed in the calculations, (c) irreducible
representations considered for each point group, and (d) for each
irreducible representation, the size of the CI expansion ($N_{total}$)
for each cluster are depicted. From Table \ref{tab:ci-calc-details}
it is obvious that most of the CI calculations were of the FCI type,
which are exact for the chosen set of active MOs. Furthermore, in
the calculations in which approaches such as QCI or MRSDCI were used,
the size of the CI expansion is quite large. This means that the CI
calculations performed in this work are fairly large scale, indicating
that the computed optical absorption spectra are numerically accurate.
Next, for these clusters, we discuss in detail the calculated ground
state geometries, followed by their optical absorption spectra.

\begin{table}
	\caption{For each cluster, the type of CI approach used for the calculations
		of the optical properties, point group symmetry employed during the
		CI calculations, and the total number of configurations ($N_{total}$)
		in the calculation are listed below. The value of $N_{total}$ corresponds
		to aug-cc-pVTZ basis set based CI calculations. \label{tab:ci-calc-details}}
	
	\centering{}%
	\begin{tabular}{|c|c|c|c|c|c|}
		\hline 
		{\scriptsize{}Cluster} & {\scriptsize{}Structure} & {\scriptsize{}Method} & {\scriptsize{}Point group} & {\scriptsize{}Symmetry} & {\scriptsize{}$N_{total}$}\tabularnewline
		&  &  & {\scriptsize{}used} &  & \tabularnewline
		\hline 
		{\scriptsize{}Li$_{2}$} & {\scriptsize{}Linear} & {\scriptsize{}FCI} & {\scriptsize{}C$_{1}$} & {\scriptsize{}$^{1}$A} & {\scriptsize{}5886}\tabularnewline
		\hline 
		{\scriptsize{}Li$_{3}$} & {\scriptsize{}Linear} & {\scriptsize{}FCI} & {\scriptsize{}C$_{1}$} & {\scriptsize{}$^{2}$A} & {\scriptsize{}575960}\tabularnewline
		\hline 
		{\scriptsize{}Li$_{3}$} & {\scriptsize{}Isosceles triangle} & {\scriptsize{}FCI} & {\scriptsize{}C$_{2v}$} & {\scriptsize{}$^{2}$A$_{1}$} & {\scriptsize{}137956}\tabularnewline
		&  &  &  & {\scriptsize{}$^{2}$B$_{1}$} & {\scriptsize{}129520}\tabularnewline
		&  &  &  & {\scriptsize{}$^{2}$B$_{2}$} & {\scriptsize{}137396}\tabularnewline
		\hline 
		{\scriptsize{}Li$_{4}$} & {\scriptsize{}Rhombus} & {\scriptsize{}FCI} & {\scriptsize{}D$_{2h}$} & {\scriptsize{}$^{1}$A$_{g}$} & {\scriptsize{}1853578}\tabularnewline
		&  &  &  & {\scriptsize{}$^{1}$B$_{1u}$} & {\scriptsize{}1846246}\tabularnewline
		&  &  &  & {\scriptsize{}$^{1}$B$_{2u}$} & {\scriptsize{}1844485}\tabularnewline
		&  &  &  & {\scriptsize{}$^{1}$B$_{3u}$} & {\scriptsize{}1802190}\tabularnewline
		\hline 
		{\scriptsize{}Be$_{2}^{+}$} & {\scriptsize{}Linear} & {\scriptsize{}FCI} & {\scriptsize{}C$_{1}$} & {\scriptsize{}$^{2}$A} & {\scriptsize{}419868}\tabularnewline
		\hline 
		{\scriptsize{}Be$_{3}^{+}$} & {\scriptsize{}Linear} & {\scriptsize{}QCI} & {\scriptsize{}D$_{2h}$} & {\scriptsize{}$^{2}$A$_{g}$} & {\scriptsize{}7393226}\tabularnewline
		&  &  &  & {\scriptsize{}$^{2}$B$_{1u}$} & {\scriptsize{}7393210}\tabularnewline
		&  &  &  & {\scriptsize{}$^{2}$B$_{2u}$} & {\scriptsize{}7286869}\tabularnewline
		&  &  &  & {\scriptsize{}$^{2}$B$_{3u}$} & {\scriptsize{}7286869}\tabularnewline
		\hline 
		{\scriptsize{}B$_{2}^{+}$} & {\scriptsize{}Linear} & {\scriptsize{}FCI} & {\scriptsize{}D$_{2h}$} & {\scriptsize{}$^{2}$A$_{g}$} & {\scriptsize{}6365216}\tabularnewline
		&  &  &  & {\scriptsize{}$^{2}$B$_{1u}$} & {\scriptsize{}6365216}\tabularnewline
		&  &  &  & {\scriptsize{}$^{2}$B$_{2u}$} & {\scriptsize{}6323328}\tabularnewline
		&  &  &  & {\scriptsize{}$^{2}$B$_{3u}$} & {\scriptsize{}6323328}\tabularnewline
		\hline 
		{\scriptsize{}B$_{3}^{+}$} & {\scriptsize{}Equilateral triangle} & {\scriptsize{}MRSDCI} & {\scriptsize{}C$_{1}$} & {\scriptsize{}$^{1}$A} & {\scriptsize{}8422193}\tabularnewline
		\hline 
	\end{tabular}
\end{table}

\subsection{Geometry}

\begin{figure}[H]
	\includegraphics[scale=0.5]{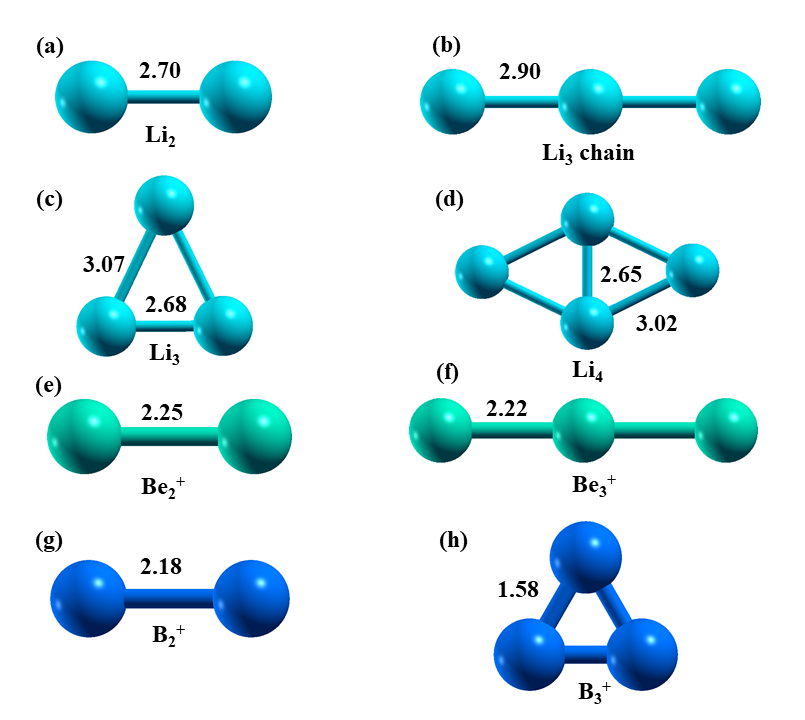}
	
	\caption{Optimized geometry of (a) Li$_2$, (b) Li$_3$ linear, (c) Li$_3$ isosceles triangular, (d) Li$_4$, (e) Be$_2^+$, (f) Be$_3^{+}$ linear, (g) B$_2^+$, and (h) B$_3^{+}$ equilateral triangular clusters considered in this work.
		The geometry optimization has been performed using the CCSD method,
		and aug-cc-pVTZ basis sets. All the listed bond lengths are in Å units.
		\label{fig:all_clusters-geo}}
\end{figure}

The simplest cluster of lithium is lithium dimer with the D$_{\infty h}$
point group symmetry. We obtained the optimized bond length of Li$_{2}$
cluster to be 2.70 $\textrm{Å}$, as shown in Fig. \ref{fig:all_clusters-geo}(a).
This result is in excellent agreement with the bond length 2.68 $\textrm{Å}$
reported by Wheeler\emph{ et al.} \cite{Wheeler_et_al}, who performed
the calculations at the CCSD/CCSD(T) level of theory using the Dunning
correlation-consistent polarized core-valence triple/quadruple-zeta
cc-pwCVXZ basis sets. Florez \emph{et al.} \cite{Florez_et_al} performed
density functional theory (DFT) calculations using the B3LYP and BLYP
functionals, and reported the bond lengths to be 2.70 $\textrm{Å}$,
and 2.71 $\textrm{Å}$, respectively, again in excellent agreement
with our result. Furthermore, our calculated bond length is also in
a very good agreement with the experimentally measured value 2.67
$\textrm{Å}$, reported by Huber\cite{Huber}. 

As far as Li$_{3}$ cluster is concerned, two isomers namely linear
and isosceles triangle were found to be stable. The equilateral triangular
structure of Li$_{3}$ cluster is not stable, and undergoes Jahn-Teller
distortion to acquire the isosceles triangular structure. The linear
structure has the D$_{\infty h}$ point-group symmetry, with the optimized
equal bond lengths of 2.90 $\textrm{Å}$ (see Fig. \ref{fig:all_clusters-geo}(b)),
in excellent agreement with the value 2.89 $\textrm{Å}$, reported
by Jones \emph{et al.}\cite{Jones_et_al}. The lowest-energy geometry
of the Li$_{3}$ cluster is an isosceles triangle with the C$_{2v}$
point-group symmetry. The CCSD-level optimized bond lengths for this
structure are found to be 2.68 and 3.07 Å, with the bond angles 51.73$^{\circ}$
and 64.13$^{\circ}$(see Fig. \ref{fig:all_clusters-geo}(c)). We
note that by performing DFT calculations, Jones \emph{et al.} \cite{Jones_et_al}
obtained the bond lengths of 2.82 and 3.37 Å, that are significantly
different as compared to our results.

The lowest-energy structure for the Li$_{4}$ cluster has a rhombus
shape, with D$_{2h}$ point group\cite{Jones_et_al,Wheeler_et_al},
as shown in Fig. \ref{fig:all_clusters-geo}(d). Our optimized bond
lengths of the side and minor diagonal of the rhombus structure are
3.02 and 2.65 Å, respectively, which are in excellent agreement with
the values 3.04 and 2.62 Å reported by Jones \emph{et al.} \cite{Jones_et_al}. 

The optimized bond length of the Be$_{2}^{+}$ cluster with the D$_{\infty h}$
point group is found to be 2.25 Å (see Fig. \ref{fig:all_clusters-geo}(e)),
in good agreement with the reported bond length 2.21 Å, obtained from
DFT calculations by Srinivas \emph{et al.} \cite{Srinivas_et_al}.
Our lowest-energy optimized structure of Be$_{3}^{+}$ cluster also
has a linear geometry, with two equal bond lengths 2.22 Å, as shown
in Fig. \ref{fig:all_clusters-geo}(f). This value of the bond length
is in very good agreement with the value 2.19 Å, computed by Srinivas
\emph{et al.} \cite{Srinivas_et_al} using DFT. 

As far as B$_{2}^{+}$ cluster is concerned, we computed its minimum-energy
bond length to be 2.18 Å (see Fig. \ref{fig:all_clusters-geo}(g)),
which is 0.18 Å larger than the value 2 Å reported by Hanley \emph{et
al.}\cite{Hanley_et_al}. We attribute this difference to two factors,
namely, smaller basis set (6-31G$^{*}$), coupled with a lower-level
CI methodology used by the authors.\cite{Hanley_et_al}. Our optimized
structure of B$_{3}^{+}$ cluster is an equilateral triangle of sides
1.58 Å, with the $D_{3h}$ point-group symmetry, as shown in Fig.
\ref{fig:all_clusters-geo}(h). Hanley\emph{ et al.} \cite{Hanley_et_al}
using a CI approach, along with the 6-31G$^{*}$ basis set, also obtained
the optimized structure to be an equilateral triangle for the B$_{3}^{+}$,
but with a bond length of 1.53 Å, which is 0.05 Å smaller than our
result. We again attribute the differences to the choice of a smaller
basis set, coupled with a lower-level correlation methodology as compared
to the CCSD approach used by us.

\subsection{Peak locations}

\begin{figure}[H]
	\includegraphics[scale=0.7]{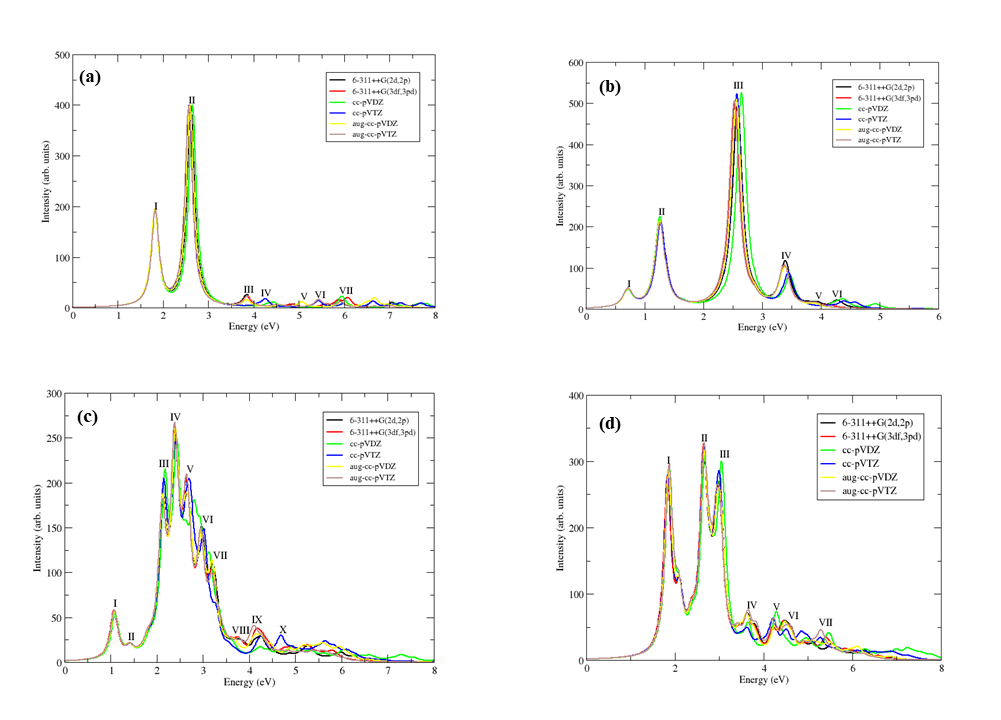}\caption{Optical absorption spectra of (a) Li$_{2}$, (b) Li$_{3}$ linear, (c) Li$_{3}$ triangular, and (d) Li$_{4}$ clusters computed using various
		basis sets and the frozen core FCI method. The uniform line-width
		0.1 eV is used to plot the spectrum. \label{fig:li_clusters-spectra}}
\end{figure}

Li$_{2}$ dimer, with just two active electrons within the frozen-core
approximation, is the smallest many-electron cluster considered in
this work. Therefore, very high-quality correlated-electron calculations
using large basis sets are possible for this system, not just for
its ground states, but also for the excited states. As a result, this
case can provide us deep insights into the influence of the choice
of basis functions on the calculated excited state properties and
the photoabsorption spectra. For the calculations, we employed the
frozen-core FCI method using six basis sets of varying sizes, namely,
6-311++G(2d,2p), 6-311++G(3df,3pd), cc-pVDZ, cc-pVTZ, aug-cc-pVDZ
and aug-cc-pVTZ, and the computed spectra are presented in Fig. \ref{fig:li_clusters-spectra}(a).
All the virtual molecular orbitals generated during the RHF calculations
were used in the CI calculations, i.e., no unoccupied orbitals were
discarded. As a result, the frozen-core FCI results presented here
are the best ones possible for the chosen basis sets. 

For the Li$_{2}$ dimer, the peak locations in the computed spectra
are presented in Table S1 of the Supporting information (SI), from
which it is obvious that for the first two peaks the excitation energies
calculated using different basis sets are in very good agreement with
each other. This is encouraging because from Fig. \ref{fig:li_clusters-spectra}(a)
it is obvious that most of the oscillator strength of the absorption
spectrum is confined to these two peaks. However, starting from the
third peak onward, we start seeing differences in the excitation energies
predicted by different basis sets. For the third peak, the predicted
peak locations can be classified in two groups: (a) those predicted
by correlation-consistent basis sets cc-pVDZ and cc-pVTZ, and (b)
the ones predicted by 6-311G++ and augmented correlation consistent
(aug-cc-) class of basis sets. We note that the peak locations predicted
by the former class of basis functions have values significantly larger
than those predicted by the latter class. Another noteworthy point
is that there is very good agreement among the peak locations predicted
by the second class of basis sets. As far as the location of the fourth
peak is concerned, there is good agreement among the predictions by
6-311++G(3df,3pd) and aug-cc class of basis functions, while the remaining
three basis functions predict very different values. The case of the
fifth peak is somewhat anomalous in that the agreement among the predictions
by any of the basis sets is not good. However, for higher peaks we
note that the results from the aug-cc class of basis functions are
in good agreement with each other, while other basis functions predict
widely differing results. Hong \emph{et al. }\cite{Hong_et_al} also
performed first-principles calculations of the photoabsorption spectra
of several Li$_{n}$ clusters employing the time-dependent density-functional
theory (TDDFT) methodology, and for Li$_{2}$ their predicted locations
of the first two peaks are 1.92 eV and 2.53 eV\cite{Hong_et_al}.
On comparing these with our best values of 1.83 eV and 2.57 eV, respectively,
we note: (a) our excitation energy for peak I is about 0.09 eV smaller
than theirs, while (b) our location for peak II is about 0.04 eV larger
than theirs. We attribute these differences to different computational
methodologies adopted in the two sets of calculations, and it will
be interesting to compare the computational results with the experimental
ones, whenever they are available.

The peak locations of the photoabsorption spectra of Li$_{3}$ chain
are presented in Table S2 of SI, from which it is clear that the locations
of the first two peaks converge completely for all the basis sets,
similar to the case of dimer. The third peak is the most intense peak
of the computed spectra as shown in Fig. \ref{fig:li_clusters-spectra}(b),
whose location is in good agreement for all the basis sets except
for cc-pVDZ, which predicts higher excitation energy as compared to
the rest. From the fourth peak onward, the peak locations can be classified
in two similar group as discussed previously for the case of dimer:
the peak locations predicted from correlation-consistent basis sets
cc-pVDZ and cc-pVTZ are towards the higher energy side as compared
to all other basis sets. It can also be seen that the peak positions
corresponding to the two classes of basis sets are in good agreement
within the class. 

Next, we examine the peak locations in the photoabsorption spectra
of Li$_{3}$ triangular cluster computed using various basis sets.
We note that the peak locations corresponding to the first five peaks
are in very good agreement with each other for different basis sets
as is obvious from Table S3 of SI. This result is very encouraging
because peak IV is the most intense (MI) peak of the computed spectra
as presented in Fig. \ref{fig:li_clusters-spectra}(c), and it is
crucial for a basis set to be able to accurately describe the MI peaks.
The location of this peak is 2.43 eV computed using the aug-cc-pVTZ
basis set, which is in a decent agreement with the experimentally
detected peak at 2.58 eV by Blanc et al.\emph{\cite{blanc_et_al_1991}.}
From the sixth peak onward it was observed that the peak locations
calculated using correlation consistent basis sets (cc-pVDZ and cc-pVTZ)
do not match with the other classes of basis sets. However, the peak
locations computed using the 6-31G class and the aug-cc-pVTZ continue
to be in very good agreement with each other till peak VIII, located
near 3.8 eV. The locations of higher-energy peaks beyond peak VIII
computed using these basis sets are presented in Table S4 for the
SI.\emph{ }

The peak positions of the photoabsorption spectra of Li$_{4}$ cluster
computed using various basis sets are presented in Table S5 of SI,
while the spectra are plotted in Fig. \ref{fig:li_clusters-spectra}(d). We note
that for this cluster, the excitation energies of the first five peaks
computed using different basis sets are in very good agreement with
each other. The first three peaks are much more intense as compared
to the higher energy peaks, and in peak III there are slight differences
($\approx$0.1 eV) in the peak locations predicted by different basis
sets. The two largest basis sets (6-311++G(3df,3pd), and aug-cc-pVTZ)
predict the location of peak III at 2.93 eV, while the predictions
by the rest of the basis sets are in the range 3.03--3.08 eV. From
peak VI onward we begin to observe differences among the locations
predicted by different basis sets, with a tendency towards clustering
into different classes. However, the noteworthy point is that the
intensity corresponding to these higher energy peaks is very low.\textbf{\textcolor{green}{{}
}}As far as the comparison with the experiments is concerned, the
first three photoabsorption peaks of the Li$_{4}$ cluster located
at 1.87 eV, 2.65 eV, and 2.93 eV for aug-cc-pVTZ basis set are in
excellent agreement with the experimental measurements of Blanc et
al.\emph{\cite{blanc_et_al_1991}} who detected these peaks at 1.83
eV, 2.65 eV, and 2.93 eV, respectively. 

\begin{figure}[H]
	\includegraphics[scale=0.6]{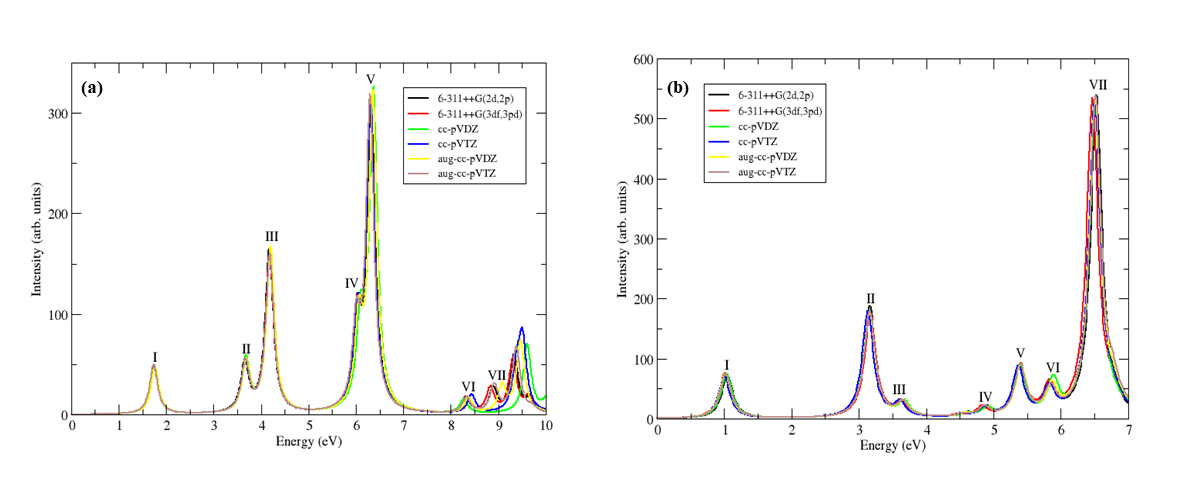}\caption{Optical absorption spectra of (a) Be$_{2}^{+}$ and (b) Be$_{3}^{+}$ clusters computed using various basis sets and frozen core FCI and QCI methods, respectively. The uniform line-width of 0.1 eV is used to plot the spectrum.\label{fig:be_clusters-spectra}}
\end{figure}

For Be$_{2}^{+}$ cluster, we present the spectra computed by different
basis sets in Fig. \ref{fig:be_clusters-spectra}(a), while the corresponding
peak locations are presented in Table S6 of SI. We note excellent
convergence of the excitation energies up to the sixth peak, beyond
which results obtained by different basis sets do not agree much with
each other. We further note that Peak V located near 6.30 eV is the
most intense peak, and, for that, the predictions of the different
basis sets are in a fairly narrow energy range 6.30-6.37 eV.

The excited-states peak locations of Be$_{3}^{+}$ cluster for different
basis sets are presented in Table S7 of SI. We notice excellent agreement
of the excited-states peak locations up to the seventh peak which
is also the most intense peak of the spectra located near 6.5 eV,
as shown in Fig. \ref{fig:be_clusters-spectra}(b). Although the peak location
of the sixth peak computed using cc-pVDZ basis set is slightly towards
the higher energy region as compared to all other basis sets, but
the difference is small. Noteworthy point is that these basis sets
are able to achieve convergence in the peak positions in Be$_{_{2}}^{+}$
and Be$_{_{3}}^{+}$ photoabsorption spectra up to much higher excitation
energies, as compared to the Li clusters. 

\begin{figure}[H]
	\includegraphics[scale=0.6]{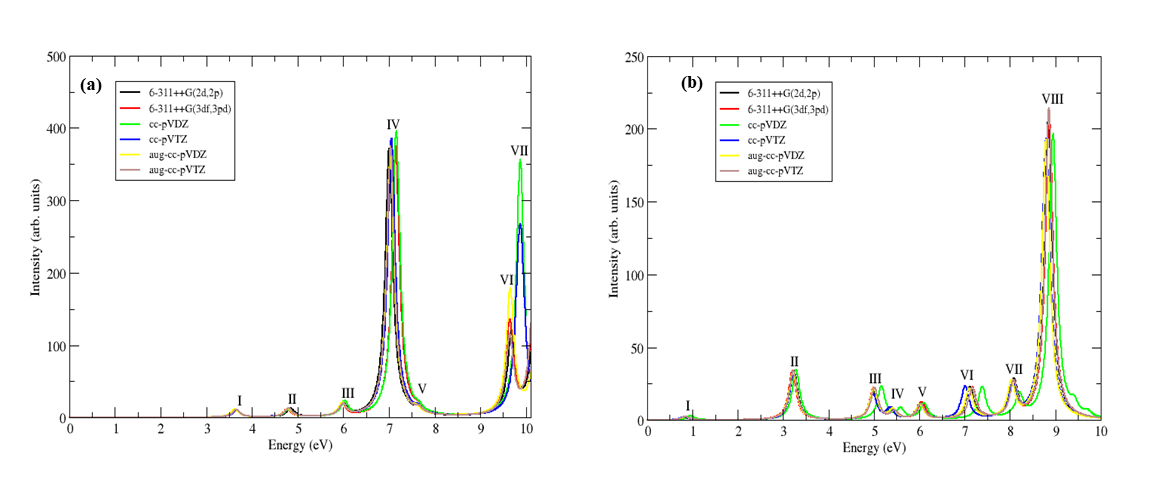}\caption{Optical absorption spectra of (a) B$_{2}^{+}$ and (b) B$_3^{+}$ cluster computed using various basis sets and frozen core FCI and MRSDCI methods, respectively. The uniform line-width 0.1 eV is used to plot the spectrum.\label{fig:b_clusters-spectra}}
\end{figure}

The excited-states peak positions corresponding to the photoabsorption
spectra of B$_{2}^{+}$ cluster are presented in Table S8 of SI. We
notice excellent agreement of the peak energies corresponding to first
three peaks for all the basis sets. The fourth peak is the most intense
peak of the spectra, as shown in Fig. \ref{fig:b_clusters-spectra}(a) for whose
location excellent agreement has been achieved for 6-311++G (2d, 2p),
cc-pVTZ, aug-cc-pVDZ, and aug-cc-pVTZ basis sets, indicating complete
convergence. However, the excitation energies for peak IV computed
using the 6-311++G (3df, 3pd) and cc-pVDZ basis sets are about 0.1
eV higher, as compared to other basis sets. As far as peak V is concerned,
which is a very weak shoulder of peak IV, we again observe excellent
convergence for all the basis sets, except cc-pVDZ which fails to
predict the peak. From the sixth peak onward, as discussed previously,
the predicted peak locations can be classified into two groups: (a)
larger basis sets of 6-311++G and aug-cc- type, and (b) smaller basis
sets cc-pVDZ and cc-pVTZ, with the peak locations predicted by individual
classes being in very good agreement with each other. 

The peak locations corresponding to the excited-states of the photoabsorption
spectra of B$_{3}^{+}$ cluster are presented in Table S9 of SI, while
the calculated spectra are plotted in Fig. \ref{fig:b_clusters-spectra}(b)
. For this cluster, we get eight well-separated peaks in the explored
energy range, with peak VIII located near 8.9 eV being the most intense.
We note that the peak energies corresponding to all the basis sets
converge excellently up to the peak VIII, except those predicted by
the cc-pVDZ basis set, which are consistently higher. We have noticed in Fig.\ref{fig:be_clusters-spectra} and Fig.\ref{fig:b_clusters-spectra}, only the deep valence excitation energies are dependent on the choice of basis sets. This behavior can be a consequence of the frozen-core approximation, which we have employed in the calculations. To verify this, we have computed the optical absorption spectra of Li$_2$ and Be$_{2}^{+}$ clusters by also including the core excitations within the large-scale QCI method. We found that the optical spectra of these clusters computed by including core excitations agrees completely with the absorption spectra computed after employing frozen-core approximation, as shown in Fig.S1 and Fig.S2 of the SI. Therefore, the frozen-core approximation does not alter the absorption spectra of small clusters. 

Based on the peak positions of the individual clusters discussed above,
we observe the following general trends: (a) peak locations for all
the clusters used in this study are in very good agreement for all
the basis sets up to the most intense peak of the spectra, except
for cc-pVDZ basis set. (b) the excited-states peak locations beyond
the most intense peak can be classified in two groups, in which the
peak locations calculated using correlation-consistent basis sets
do not match with the peak locations computed using all other basis
sets, and (c) for the cc-pVDZ basis sets peaks are located at higher
energies as compared to the rest of the basis functions.
As the basis-set dependence of the optical properties is different for the density functional theory compared to the wave function-based large-scale configuration-interaction method, it will be interesting to explore the optical properties of clusters using time-dependent density functional theory (TD-DFT) and compare it with our results. We found that the first peak of the optical absorption spectra of B$_3^+$ cluster is located at 0.84 eV when computed using large-scale MRSDCI calculations along with a large aug-cc-pVTZ basis set. However, when the calculations are performed using TD-DFT method with B3LYP functional and the same basis set, it is obtained at 0.95 eV. The most intense peak of the optical absorption spectra is located at 8.85 eV using the MRSDCI approach, which is found to be at 9.35 eV by employing the TD-DFT method. The calulated optical absorption spectra and excited-states peak locations of B$_3^+$ cluster corresponding to TD-DFT calculations are provided in the Fig.S3 and Table S10 of the SI. We also report that the variations in the peak locations of the photoabsorption spectra of B$_3^+$ computed using various basis sets and TD-DFT method are lesser than the wave function-based CI method.    

\subsection{Oscillator strength}

In addition to the excitation energy, the next important quantity
determining the profile of the absorption spectrum is the oscillator
strength ($f$) corresponding to various optical transitions, connecting
the ground state to the excited state in question. The oscillator
strength calculated using Eq. \ref{eq:osc-strength} is determined
by the excitation energy of the state involved, and the corresponding
transition dipole moment (TDM). The TDM being a matrix element, is,
in turn, determined by the many-particle wave functions of the ground
and the excited state that it connects. Another important quantity
is the polarization of the photon involved in a given optical transition
(peak), which can be measured in oriented samples. The polarization
is a consequence of the point-group symmetry of the concerned molecule,
and hence should be independent of the basis set employed. In this
section, we discuss the convergence of the oscillator strengths and
photon polarizations associated with various peaks of the calculated
spectra. In Table \ref{tab:osc-strngth-li2}, we present the oscillator strengths
corresponding to the first peak, and the most intense peak (peak II)
of the spectra of Li$_{2}$ computed using different basis functions.
Additionally, the table also contains the dominant configurations
contributing to the many-particle wave functions of the excited states
involved.

\begin{table}
	\caption{Comparison of oscillator strengths and the dominant configurations
		contributing to the many-particle wave functions for peaks I and II
		of Li$_{2}$ cluster calculated using different basis sets. In the
		“Polarization” column, $\parallel$ indicates photon polarization
		along the direction of the molecule (longitudinal polarization), while
		$\perp$ indicates polarization perpendicular to the molecular axis
		(transverse polarization). Note that the transversely polarized states
		are doubly degenerate, therefore, the oscillator strength corresponding
		to those is the sum of both the contributions. ’H’ and ’L’ stand for
		HOMO and LUMO orbitals. \label{tab:osc-strngth-li2}}
	
	\centering{}{\small{}}%
	\begin{tabular}{|c|c|c|c|c|c|c|}
		\hline 
		{\scriptsize{}Basis Set} & \multicolumn{1}{c}{} & \multicolumn{1}{c}{{\scriptsize{}Peak I}} &  & \multicolumn{1}{c}{} & \multicolumn{1}{c}{{\scriptsize{}Peak II}} & \tabularnewline
		\cline{2-7} \cline{3-7} \cline{4-7} \cline{5-7} \cline{6-7} \cline{7-7} 
		& {\scriptsize{}Polarization} & {\scriptsize{}$f$} & {\scriptsize{}Configurations} & {\scriptsize{}Polarization} & {\scriptsize{}$f$} & {\scriptsize{}Configurations}\tabularnewline
		\hline 
		\hline 
		{\scriptsize{}6-311++G(2d,2p)} & {\scriptsize{}$\parallel$} & {\scriptsize{}0.460} & {\scriptsize{}$\arrowvert H\rightarrow L\rangle$} & {\scriptsize{}$\perp$} & {\scriptsize{}0.971} & {\scriptsize{}$\arrowvert H\rightarrow L+2\rangle$}\tabularnewline
		&  &  & {\scriptsize{}$\arrowvert H\rightarrow L+3\rangle$} &  &  & {\scriptsize{}$\arrowvert H\rightarrow L+7\rangle$}\tabularnewline
		\hline 
		{\scriptsize{}6-311++G(3df,3pd)} & {\scriptsize{}$\parallel$} & {\scriptsize{}0.456} & {\scriptsize{}$\arrowvert H\rightarrow L\rangle$} & {\scriptsize{}$\perp$} & {\scriptsize{}0.966} & {\scriptsize{}$\arrowvert H\rightarrow L+2\rangle$}\tabularnewline
		&  &  & {\scriptsize{}$\arrowvert H\rightarrow L+3\rangle$} &  &  & {\scriptsize{}$\arrowvert H\rightarrow L+7\rangle$}\tabularnewline
		\hline 
		{\scriptsize{}cc-PVDZ} & {\scriptsize{}$\parallel$} & {\scriptsize{}0.463} & {\scriptsize{}$\arrowvert H\rightarrow L\rangle$} & {\scriptsize{}$\perp$} & {\scriptsize{}0.970} & {\scriptsize{}$\arrowvert H\rightarrow L+1\rangle$}\tabularnewline
		&  &  & {\scriptsize{}$\arrowvert H\rightarrow L;H\rightarrow L+2\rangle$} &  &  & {\scriptsize{}$\arrowvert H\rightarrow L+1;H\rightarrow L+5\rangle$}\tabularnewline
		\hline 
		{\scriptsize{}cc-pVTZ} & {\scriptsize{}$\parallel$} & {\scriptsize{}0.455} & {\scriptsize{}$\arrowvert H\rightarrow L\rangle$} & {\scriptsize{}$\perp$} & {\scriptsize{}0.969} & {\scriptsize{}$\arrowvert H\rightarrow L+1\rangle$}\tabularnewline
		&  &  & {\scriptsize{}$\arrowvert H\rightarrow L+4\rangle$} &  &  & {\scriptsize{}$\arrowvert H\rightarrow L+6\rangle$}\tabularnewline
		\hline 
		{\scriptsize{}aug-cc-pVDZ} & {\scriptsize{}$\parallel$} & {\scriptsize{}0.462} & {\scriptsize{}$\arrowvert H\rightarrow L\rangle$} & {\scriptsize{}$\perp$} & {\scriptsize{}0.972} & {\scriptsize{}$\arrowvert H\rightarrow L+1\rangle$}\tabularnewline
		&  &  & {\scriptsize{}$\arrowvert H\rightarrow L+3\rangle$} &  &  & {\scriptsize{}$\arrowvert H\rightarrow L+6\rangle$}\tabularnewline
		\hline 
		{\scriptsize{}aug-cc-pVTZ} & {\scriptsize{}$\parallel$} & {\scriptsize{}0.454} & {\scriptsize{}$\arrowvert H\rightarrow L\rangle$} & {\scriptsize{}$\perp$} & {\scriptsize{}0.966} & {\scriptsize{}$\arrowvert H\rightarrow L+2\rangle$}\tabularnewline
		&  &  & {\scriptsize{}$\arrowvert H\rightarrow L+3\rangle$} &  &  & {\scriptsize{}$\arrowvert H\rightarrow L+7\rangle$}\tabularnewline
		\hline 
	\end{tabular}{\small\par}
\end{table}

\textcolor{green}{{} }From Table \ref{tab:osc-strngth-li2}
it is obvious that the oscillator strengths computed using various
basis functions for both the peaks are in very good agreement with
each other. We also note that the direction of the polarization of
the photons involved in a given optical transitions are of the excited-states
corresponding to the first and second peak of the spectra of Li$_{2}$
cluster are parallel and perpendicular to molecular axis, respectively,
irrespective of the basis set. 

The oscillator strengths corresponding to the first and most intense
peaks of the spectra of Li$_{3}$ chain and triangular clusters are
presented in Table S11 and Table S12,
respectively. We note that the oscillator strengths of the first peaks
both of Li$_{3}$ chain, and the triangular cluster, computed using
the different basis sets are in excellent agreement with each other.
The oscillator strengths corresponding to the most intense peaks of
Li$_{3}$, i.e. peak III of the chain and peak IV for the triangular
cluster, calculated using various basis sets can be classified into
two groups: (a) those calculated using correlation-consistent (cc-pVTZ
and cc-pVDZ) class of basis sets, and (b) those computed using 6-33++G-
and augmented correlation-consistent (aug-cc-) class of basis sets.
The oscillator strength calculated by the first class of basis sets
is comparatively higher than the second class of basis sets. But,
the relative maximum difference between oscillator strengths of different
classes is close to 6\%, which is fairly acceptable.

The oscillator strengths corresponding to the first and the most intense
peak (peak II) in the photoabsorption spectra of Li$_{4}$ cluster
are presented in Table S13. We note that the oscillator strengths of peak I are in good agreement with each other for all the basis sets except for cc-pVDZ and aug-cc-pVTZ. For these basis sets the oscillator strength is comparatively larger.
For peak II, we note that the difference in the oscillator strength
computed by cc-pVDZ and 6-311++G (3df, 3pd) basis sets is about 5\%,
which is again quite small. 

We present the oscillator strengths corresponding to the first and
the most intense peaks of the cationic beryllium clusters Be$_{2}^{+}$
and Be$_{3}^{+}$ in tables S14, and S15,
respectively. We note very good agreement on the oscillator strengths
of both the peaks of the Be$_{2}^{+}$ and Be$_{3}^{+}$ clusters
for all the basis sets. The maximum relative disagreement we find
among the oscillator strengths for a given peak is around 6\%.

Finally, we discuss the oscillator strengths of the first and the
most intense peaks of the B$_{2}^{+}$ and B$_{3}^{+}$ clusters presented
in tables S16, and S17, respectively. We note that both for B$_{2}^{+}$ and B$_{3}^{+}$
clusters, the oscillator strengths of the first peaks are two orders
of magnitude smaller than those of their most intense peaks, indicating
that the first peaks for both the clusters are relatively feeble.
Nevertheless, the oscillator strengths of the first peaks of the photoabsorption
spectra of the two clusters calculated using various basis sets are
in very good agreement with each other. As far as the most intense
peaks are concerned, both for B$_{2}^{+}$ and B$_{3}^{+}$ we see
the following pattern: oscillator strengths computed using 6-311++G-
and aug-cc-pVTZ basis sets are in very good agreement with each other,
while those computed using other basis sets differ from them somewhat. 

\subsection{Wave function analysis}

Next, we examine the dominant configurations contributing to the CI
wave functions of the excited states contributing to various peaks.
The dominant configurations corresponding to the excited-states CI
wave functions of peak I and peak II of Li$_{2}$ are presented in
Table \ref{tab:osc-strngth-li2}. We note that for peak I, the main
contribution to the corresponding excited state wave function is from
the singly excited configuration $\arrowvert H\rightarrow L\rangle$
for all the basis sets. However, the next important configuration
to the same wave function depends on the class of basis set employed:
(a) it is $\arrowvert H\rightarrow L+3\rangle$ single excitation
when calculations are performed using larger basis sets of the type
6-311++ and aug-cc, but (b) for smaller basis sets, this configuration
is found to be $\arrowvert H\rightarrow L+4\rangle$ for cc-PVTZ basis,
and $\arrowvert H\rightarrow L;H\rightarrow L+2\rangle$ for the cc-PVDZ
set. Peak II is due to two degenerate excited states to which the
dominant contributions are from configurations $\arrowvert H\rightarrow L+2\rangle$
and $\arrowvert H\rightarrow L+7\rangle$, for the calculations performed
using 6-31G++ and aug-cc-PVTZ type basis sets. But, for the calculations
performed with smaller basis sets, the dominant configurations is
$\arrowvert H\rightarrow L+1\rangle$, while the next important configuration
can be $\arrowvert H\rightarrow L+6\rangle$ or $\arrowvert H\rightarrow L+1;H\rightarrow L+5\rangle$,
depending on the basis set. Thus, we can draw the following general
conclusion regarding this: (a) for large basis set calculations, for
a given peak, the configurations are in perfect agreement with each
other, and (b) the configurations predicted by calculations performed
using smaller basis sets such as cc-PVDZ are found to be different
as compared to those obtained in larger basis set calculations. 

The dominant configurations for the wave functions corresponding to
peak I and the most intense peak (peak III) of Li$_{3}$ chain, computed
using various basis sets are presented in Table S11.
We find that for the first peak the dominant configuration is $\arrowvert H-1\rightarrow H\rangle$
for all the basis sets except for aug-cc-pVTZ. For the aug-cc-pVTZ
the dominant configurations contributing to the excited state wave
function are different compared to other basis sets, because of the
reversal of ground and excited states. However, because the peak energies
and oscillator strength for the state are in excellent agreement with
all other basis sets implies that we have obtained correct quantitative
description of the excited states even with this basis set. For peak
III the main contribution to the excited state wave function is from
$\arrowvert H-1\rightarrow L+2\rangle$ for the larger 6-311++ and
aug-cc class of basis sets, while it is from $\arrowvert H-1\rightarrow L+1\rangle$
for the smaller cc-pVTZ and cc-pVDZ basis sets. 

The main configurations contributing to the excited states wave functions
of peak I and the most intense peak (peak IV) of Li$_{3}$ triangular
cluster, computed using various basis sets are presented in Table S12. We note that for the first peak,
the main contribution to the wave function is from configurations
$\arrowvert H\rightarrow L+14\rangle$ or $\arrowvert H\rightarrow L+13\rangle$
for the larger 6-311++G and aug-cc class of basis sets. For the correlation-consistent
basis sets (cc-pVDZ and cc-pVTZ) the main contribution is due to the
configuration $\arrowvert H\rightarrow L+2\rangle$. For the fourth
peak, the dominant configuration is $\arrowvert H-1\rightarrow L\rangle$,
irrespective of the type of basis set used for the calculation. 

The dominant configurations corresponding to the excited states wave
functions of peak I and the most intense peak of the spectra (peak
II) of the Li$_{4}$ cluster are presented in Table S13.
For the first peak, the main contribution is from the configuration
$\arrowvert H\rightarrow L+1\rangle$ for all the basis sets, while
for peak II it is $\arrowvert H-1\rightarrow L\rangle$ for all the
basis sets. Thus, we have excellent agreement among all the basis
sets when it comes to the most important configuration for both the
peaks of the Li$_{4}$ cluster.

The important configurations corresponding to the excited states wave
function of peak I and the most intense peak (peak V) of the photoabsorption
spectra of Be$_{2}^{+}$ cluster are presented in Table S14.
The dominant configuration contributing to peak I is $\arrowvert H\rightarrow L\rangle$
for the 6-311++G class of basis sets, and $\arrowvert H\rightarrow L+2\rangle$
for the rest. For peak V, the main configuration contributing to the
CI wave function is $\arrowvert H-1\rightarrow L+1\rangle$ for 6-311++G
class of basis sets, and $\arrowvert H-1\rightarrow L\rangle$ for
the rest of the sets. 

The configurations dominating the excited state CI wave functions
of peak I and the most intense peak (peak VII) of Be$_{3}^{+}$ cluster
are listed in Table S15. We note that the most important configurations contributing to peak I can be classified
in two groups: (a) for larger 6-311++G(3df,3pd) and aug-cc class of
basis sets the dominant configuration is $\arrowvert H\rightarrow L+1\rangle$,
(b) while for smaller basis sets dominant configuration is highly
basis set dependent. For peak VII, the doubly-excited configurations
$\arrowvert H-2\rightarrow L;H\rightarrow L+2\rangle$ and $\arrowvert H-1\rightarrow L;H\rightarrow L+4\rangle$
dominate the excited-state wave functions for the larger 6-311++G(3df,3pd)
and aug-cc class of basis sets, but vary significantly for the rest.

Most important configurations contributing to the wave functions for
peak I and the most intense peak (peak IV) of B$_{2}^{+}$ cluster
are presented in Table S16. It is obvious
that the double-excitation $\arrowvert H-1\rightarrow L;H\rightarrow L\rangle$
contributes the most to peak I for all the basis sets. The dominant
configurations contributing to the wave functions of peak IV are $\arrowvert H\rightarrow L+5\rangle$
and $\arrowvert H\rightarrow L+11\rangle$ for all the basis sets
except the cc-pVDZ/cc-pVTZ, for which instead of $\arrowvert H\rightarrow L+11\rangle$,
the double excitation $\arrowvert H-1\rightarrow L;H\rightarrow L\rangle$
contributes.

Finally, we present the dominant configurations in the CI wave functions
corresponding to peak I, and the most intense peak (peak VIII), of
B$_{3}^{+}$ cluster in Table S17. The configuration
with maximum contribution to the excited state wave functions for
peak I is $\arrowvert H\rightarrow L\rangle$, irrespective of the
basis set. The next dominant configuration is basis-set dependent,
however, it is a double excitation in all the cases. The dominant
configuration corresponding to the CI wave function of peak VIII is
the double excitation $\arrowvert H-1\rightarrow L;H\rightarrow L+3\rangle$
for all the basis sets. 

The detailed wave function analysis for all the peaks of the optical
absorption spectra of clusters considered in this work using the largest
aug-cc-pVTZ basis set is provided in Table S18-S25 of the SI.

\section{Conclusion\label{sec:conclusion}}

In this work, we presented electron-correlated calculations of the
optical absorption spectra of small neutral and ionic clusters using
various basis sets. First, the stable geometries of various clusters
were determined at the CCSD level of theory, using the aug-cc-PVTZ
basis set. For the ground and the excited state wave functions calculations
needed to compute the absorption spectra, we used the FCI, QCI, and
MRSDCI approaches depending upon the size of the clusters. The CI
calculations were performed using six different basis sets, namely,
6-311++G(2d,2p), 6-311++G(3df,3pd), cc-pVDZ , cc-pVTZ, aug-cc-pVDZ,
and aug-cc-pVTZ. 

We observed that the optical absorption spectra of all these clusters
exhibit excellent convergence for all the basis sets in the lower
energy range. However, usually after the first two peaks, the shift
in peak locations for cc-pVDZ and cc-pVTZ basis set are noted in all
likelihood because of the lack of diffuse basis functions in these
sets. If we use augmented basis sets, the absorption spectra show
good agreement with the results computed using other similar basis
sets. Although aug-cc-pVDZ basis set has a relatively smaller number
of basis functions as compared to aug-cc-pVTZ basis set, the agreement
between the spectra computed using the two basis sets is very good.
Because the number of two-electron integrals increases as $N^{4}$
where $N$ is the number of basis functions in basis set, we can reduce
the computational cost significantly by using aug-cc-pVDZ basis set
instead of larger Pople's basis sets, and aug-cc-pVTZ basis set. Thus,
our general recommendation is that for optical absorption calculations
one should use a basis set containing diffuse functions, i.e., of
the aug-cc- type. However, whether one should use aug-cc-pVDZ, or
a larger set, should be decided by the available computational resources. 

We believe that the CI calculations presented in this work are quite
accurate, as is obvious from the fact that our obtained results are
in very good agreement with the experiments for Li$_{3}$ and Li$_{4}$
clusters. Therefore, it will be of interest to compare our results
on other clusters also with the experiments, as and when they are
performed.

\section*{Associated Content}
\subsection*{Supporting Information}

In the supporting information file, we have provided the peak locations, oscillator strengths, and dominant excited state configurations corresponding to the optical absorption spectra of all the clusters for all the
basis sets considered in this work. The SI file also contains the
details of the many-particle wave functions of excited states contributing
to the peaks in the optical absorption spectrum of clusters for aug-cc-pVTZ
basis set.

\section*{Author Information}

\subsection*{Corresponding Author}

Alok Shukla: Department of Physics, Indian Institute of Technology Bombay, Powai,
Mumbai 400076, India;   {*}E-mail: shukla@phy.iitb.ac.in

\section{Acknowledgment}

This work was supported by senior research fellowship (DST-Inspire)
provided by department of science and technology, India.

\section*{Authors}

Vikram Mahamiya: Department of Physics, Indian Institute of Technology Bombay, Powai,
Mumbai 400076, India; E-mail: mahamiyavikram@gmail.com

Pritam Bhattacharyya: Institute for Theoretical Solid State Physics, Leibniz
IFW Dresden, Helmholtzstr. 20, 01069 Dresden, Germany; E-mail: pritambhattacharyya01@gmail.com 

\section*{Notes}

The authors declare no competing financial interests.

\newpage{}

\bibliographystyle{achemso}
\addcontentsline{toc}{section}{\refname}\bibliography{small_clus}

\providecommand{\latin}[1]{#1}
\makeatletter
\providecommand{\doi}
  {\begingroup\let\do\@makeother\dospecials
  \catcode`\{=1 \catcode`\}=2 \doi@aux}
\providecommand{\doi@aux}[1]{\endgroup\texttt{#1}}
\makeatother
\providecommand*\mcitethebibliography{\thebibliography}
\csname @ifundefined\endcsname{endmcitethebibliography}
  {\let\endmcitethebibliography\endthebibliography}{}
\begin{mcitethebibliography}{52}
\providecommand*\natexlab[1]{#1}
\providecommand*\mciteSetBstSublistMode[1]{}
\providecommand*\mciteSetBstMaxWidthForm[2]{}
\providecommand*\mciteBstWouldAddEndPuncttrue
  {\def\EndOfBibitem{\unskip.}}
\providecommand*\mciteBstWouldAddEndPunctfalse
  {\let\EndOfBibitem\relax}
\providecommand*\mciteSetBstMidEndSepPunct[3]{}
\providecommand*\mciteSetBstSublistLabelBeginEnd[3]{}
\providecommand*\EndOfBibitem{}
\mciteSetBstSublistMode{f}
\mciteSetBstMaxWidthForm{subitem}{(\alph{mcitesubitemcount})}
\mciteSetBstSublistLabelBeginEnd
  {\mcitemaxwidthsubitemform\space}
  {\relax}
  {\relax}

\bibitem[{Boys}(1950)]{boys-gaussian}
{Boys},~S.~F. {Electronic Wave Functions. I. A General Method of Calculation
  for the Stationary States of Any Molecular System}. \emph{Proceedings of the
  Royal Society of London Series A} \textbf{1950}, \emph{200}, 542--554\relax
\mciteBstWouldAddEndPuncttrue
\mciteSetBstMidEndSepPunct{\mcitedefaultmidpunct}
{\mcitedefaultendpunct}{\mcitedefaultseppunct}\relax
\EndOfBibitem
\bibitem[McMurchie and Davidson(1978)McMurchie, and
  Davidson]{mcmurchie-davidson}
McMurchie,~L.~E.; Davidson,~E.~R. One- and two-electron integrals over
  cartesian gaussian functions. \emph{Journal of Computational Physics}
  \textbf{1978}, \emph{26}, 218 -- 231\relax
\mciteBstWouldAddEndPuncttrue
\mciteSetBstMidEndSepPunct{\mcitedefaultmidpunct}
{\mcitedefaultendpunct}{\mcitedefaultseppunct}\relax
\EndOfBibitem
\bibitem[Davidson and Feller(1986)Davidson, and
  Feller]{basis-set-review-davidson-feller}
Davidson,~E.~R.; Feller,~D. Basis set selection for molecular calculations.
  \emph{Chemical Reviews} \textbf{1986}, \emph{86}, 681--696\relax
\mciteBstWouldAddEndPuncttrue
\mciteSetBstMidEndSepPunct{\mcitedefaultmidpunct}
{\mcitedefaultendpunct}{\mcitedefaultseppunct}\relax
\EndOfBibitem
\bibitem[Huzinaga(1985)]{huzinaga-basis-set-review}
Huzinaga,~S. Basis sets for molecular calculations. \emph{Computer Physics
  Reports} \textbf{1985}, \emph{2}, 281 -- 339\relax
\mciteBstWouldAddEndPuncttrue
\mciteSetBstMidEndSepPunct{\mcitedefaultmidpunct}
{\mcitedefaultendpunct}{\mcitedefaultseppunct}\relax
\EndOfBibitem
\bibitem[Huzinaga(1965)]{huzinaga1}
Huzinaga,~S. Gaussian‐Type Functions for Polyatomic Systems. I. \emph{The
  Journal of Chemical Physics} \textbf{1965}, \emph{42}, 1293--1302\relax
\mciteBstWouldAddEndPuncttrue
\mciteSetBstMidEndSepPunct{\mcitedefaultmidpunct}
{\mcitedefaultendpunct}{\mcitedefaultseppunct}\relax
\EndOfBibitem
\bibitem[Ruedenberg \latin{et~al.}(1973)Ruedenberg, Raffenetti, and
  Bardo]{even-tempered1}
Ruedenberg,~K.; Raffenetti,~R.~C.; Bardo,~R.~D. \emph{Structure and Reactivity,
  Proceedings of the 1972 Boulder Conference}; Wiley: New York, 1973\relax
\mciteBstWouldAddEndPuncttrue
\mciteSetBstMidEndSepPunct{\mcitedefaultmidpunct}
{\mcitedefaultendpunct}{\mcitedefaultseppunct}\relax
\EndOfBibitem
\bibitem[Bardo and Ruedenberg(1974)Bardo, and Ruedenberg]{even-tempered2}
Bardo,~R.~D.; Ruedenberg,~K. Even‐tempered atomic orbitals. VI. Optimal
  orbital exponents and optimal contractions of Gaussian primitives for
  hydrogen, carbon, and oxygen in molecules. \emph{The Journal of Chemical
  Physics} \textbf{1974}, \emph{60}, 918--931\relax
\mciteBstWouldAddEndPuncttrue
\mciteSetBstMidEndSepPunct{\mcitedefaultmidpunct}
{\mcitedefaultendpunct}{\mcitedefaultseppunct}\relax
\EndOfBibitem
\bibitem[Huzinaga \latin{et~al.}(1985)Huzinaga, Klobukowski, and
  Tatewaki]{huzinaga-wt}
Huzinaga,~S.; Klobukowski,~M.; Tatewaki,~H. The well-tempered GTF basis sets
  and their applications in the SCF calculations on N2, CO, Na2, and P2.
  \emph{Canadian Journal of Chemistry} \textbf{1985}, \emph{63},
  1812--1828\relax
\mciteBstWouldAddEndPuncttrue
\mciteSetBstMidEndSepPunct{\mcitedefaultmidpunct}
{\mcitedefaultendpunct}{\mcitedefaultseppunct}\relax
\EndOfBibitem
\bibitem[Tatewaki and Huzinaga(1979)Tatewaki, and
  Huzinaga]{tatewaki-huzinaga-1979}
Tatewaki,~H.; Huzinaga,~S. A systematic preparation of new contracted Gaussian
  type orbital set. I. Transition metal atoms from Sc to Zn. \emph{The Journal
  of Chemical Physics} \textbf{1979}, \emph{71}, 4339--4348\relax
\mciteBstWouldAddEndPuncttrue
\mciteSetBstMidEndSepPunct{\mcitedefaultmidpunct}
{\mcitedefaultendpunct}{\mcitedefaultseppunct}\relax
\EndOfBibitem
\bibitem[Tatewaki and Huzinaga(1980)Tatewaki, and
  Huzinaga]{tatewaki-huzinaga-1980}
Tatewaki,~H.; Huzinaga,~S. A systematic preparation of new contracted
  Gaussian-type orbital sets. III. Second-row atoms from Li through ne.
  \emph{Journal of Computational Chemistry} \textbf{1980}, \emph{1},
  205--228\relax
\mciteBstWouldAddEndPuncttrue
\mciteSetBstMidEndSepPunct{\mcitedefaultmidpunct}
{\mcitedefaultendpunct}{\mcitedefaultseppunct}\relax
\EndOfBibitem
\bibitem[Hehre \latin{et~al.}(1969)Hehre, Stewart, and Pople]{pople-sto-ng}
Hehre,~W.~J.; Stewart,~R.~F.; Pople,~J.~A. Self‐Consistent
  Molecular‐Orbital Methods. I. Use of Gaussian Expansions of Slater‐Type
  Atomic Orbitals. \emph{The Journal of Chemical Physics} \textbf{1969},
  \emph{51}, 2657--2664\relax
\mciteBstWouldAddEndPuncttrue
\mciteSetBstMidEndSepPunct{\mcitedefaultmidpunct}
{\mcitedefaultendpunct}{\mcitedefaultseppunct}\relax
\EndOfBibitem
\bibitem[Ditchfield \latin{et~al.}(1971)Ditchfield, Hehre, and
  Pople]{pople-431g}
Ditchfield,~R.; Hehre,~W.~J.; Pople,~J.~A. Self‐Consistent
  Molecular‐Orbital Methods. IX. An Extended Gaussian‐Type Basis for
  Molecular‐Orbital Studies of Organic Molecules. \emph{The Journal of
  Chemical Physics} \textbf{1971}, \emph{54}, 724--728\relax
\mciteBstWouldAddEndPuncttrue
\mciteSetBstMidEndSepPunct{\mcitedefaultmidpunct}
{\mcitedefaultendpunct}{\mcitedefaultseppunct}\relax
\EndOfBibitem
\bibitem[Binkley \latin{et~al.}(1980)Binkley, Pople, and Hehre]{pople-321g}
Binkley,~J.~S.; Pople,~J.~A.; Hehre,~W.~J. Self-consistent molecular orbital
  methods. 21. Small split-valence basis sets for first-row elements.
  \emph{Journal of the American Chemical Society} \textbf{1980}, \emph{102},
  939--947\relax
\mciteBstWouldAddEndPuncttrue
\mciteSetBstMidEndSepPunct{\mcitedefaultmidpunct}
{\mcitedefaultendpunct}{\mcitedefaultseppunct}\relax
\EndOfBibitem
\bibitem[Binkley and Pople(1977)Binkley, and Pople]{pople-621g}
Binkley,~J.~S.; Pople,~J.~A. Self‐consistent molecular orbital methods. XIX.
  Split‐valence Gaussian‐type basis sets for beryllium. \emph{The Journal
  of Chemical Physics} \textbf{1977}, \emph{66}, 879--880\relax
\mciteBstWouldAddEndPuncttrue
\mciteSetBstMidEndSepPunct{\mcitedefaultmidpunct}
{\mcitedefaultendpunct}{\mcitedefaultseppunct}\relax
\EndOfBibitem
\bibitem[Krishnan \latin{et~al.}(1980)Krishnan, Binkley, Seeger, and
  Pople]{pople-6311g-6311g*}
Krishnan,~R.; Binkley,~J.~S.; Seeger,~R.; Pople,~J.~A. Self‐consistent
  molecular orbital methods. XX. A basis set for correlated wave functions.
  \emph{The Journal of Chemical Physics} \textbf{1980}, \emph{72},
  650--654\relax
\mciteBstWouldAddEndPuncttrue
\mciteSetBstMidEndSepPunct{\mcitedefaultmidpunct}
{\mcitedefaultendpunct}{\mcitedefaultseppunct}\relax
\EndOfBibitem
\bibitem[Hariharan and Pople(1973)Hariharan, and Pople]{pople1973-631g*}
Hariharan,~P.~C.; Pople,~J.~A. The influence of polarization functions on
  molecular orbital hydrogenation energies. \emph{Theoretica chimica acta}
  \textbf{1973}, \emph{28}, 213--222\relax
\mciteBstWouldAddEndPuncttrue
\mciteSetBstMidEndSepPunct{\mcitedefaultmidpunct}
{\mcitedefaultendpunct}{\mcitedefaultseppunct}\relax
\EndOfBibitem
\bibitem[Collins \latin{et~al.}(1976)Collins, von R.~Schleyer, Binkley, and
  Pople]{pople-sto3g*}
Collins,~J.~B.; von R.~Schleyer,~P.; Binkley,~J.~S.; Pople,~J.~A.
  Self‐consistent molecular orbital methods. XVII. Geometries and binding
  energies of second‐row molecules. A comparison of three basis sets.
  \emph{The Journal of Chemical Physics} \textbf{1976}, \emph{64},
  5142--5151\relax
\mciteBstWouldAddEndPuncttrue
\mciteSetBstMidEndSepPunct{\mcitedefaultmidpunct}
{\mcitedefaultendpunct}{\mcitedefaultseppunct}\relax
\EndOfBibitem
\bibitem[Francl \latin{et~al.}(1982)Francl, Pietro, Hehre, Binkley, Gordon,
  DeFrees, and Pople]{pople-631g*-second-row}
Francl,~M.~M.; Pietro,~W.~J.; Hehre,~W.~J.; Binkley,~J.~S.; Gordon,~M.~S.;
  DeFrees,~D.~J.; Pople,~J.~A. Self‐consistent molecular orbital methods.
  XXIII. A polarization‐type basis set for second‐row elements. \emph{The
  Journal of Chemical Physics} \textbf{1982}, \emph{77}, 3654--3665\relax
\mciteBstWouldAddEndPuncttrue
\mciteSetBstMidEndSepPunct{\mcitedefaultmidpunct}
{\mcitedefaultendpunct}{\mcitedefaultseppunct}\relax
\EndOfBibitem
\bibitem[Pietro \latin{et~al.}(1982)Pietro, Francl, Hehre, DeFrees, Pople, and
  Binkley]{pople-321g*}
Pietro,~W.~J.; Francl,~M.~M.; Hehre,~W.~J.; DeFrees,~D.~J.; Pople,~J.~A.;
  Binkley,~J.~S. Self-consistent molecular orbital methods. 24. Supplemented
  small split-valence basis sets for second-row elements. \emph{Journal of the
  American Chemical Society} \textbf{1982}, \emph{104}, 5039--5048\relax
\mciteBstWouldAddEndPuncttrue
\mciteSetBstMidEndSepPunct{\mcitedefaultmidpunct}
{\mcitedefaultendpunct}{\mcitedefaultseppunct}\relax
\EndOfBibitem
\bibitem[Frisch \latin{et~al.}(1984)Frisch, Pople, and
  Binkley]{pople-6311g*-larger}
Frisch,~M.~J.; Pople,~J.~A.; Binkley,~J.~S. Self‐consistent molecular orbital
  methods 25. Supplementary functions for Gaussian basis sets. \emph{The
  Journal of Chemical Physics} \textbf{1984}, \emph{80}, 3265--3269\relax
\mciteBstWouldAddEndPuncttrue
\mciteSetBstMidEndSepPunct{\mcitedefaultmidpunct}
{\mcitedefaultendpunct}{\mcitedefaultseppunct}\relax
\EndOfBibitem
\bibitem[Dunning(1989)]{DUNNING_1}
Dunning,~T.~H. Gaussian basis sets for use in correlated molecular
  calculations.I The atoms boron through neon and hydrogen. \emph{The Journal
  of Chemical Physics} \textbf{1989}, \emph{90}, 1007--1023\relax
\mciteBstWouldAddEndPuncttrue
\mciteSetBstMidEndSepPunct{\mcitedefaultmidpunct}
{\mcitedefaultendpunct}{\mcitedefaultseppunct}\relax
\EndOfBibitem
\bibitem[Kendall \latin{et~al.}(1992)Kendall, Dunning, and Harrison]{DUNNING_2}
Kendall,~R.~A.; Dunning,~T.~H.; Harrison,~R.~J. Electron affinities of the
  first-row atoms revisited.Systematic basis sets and wave functions. \emph{The
  Journal of Chemical Physics} \textbf{1992}, \emph{96}, 6796--6806\relax
\mciteBstWouldAddEndPuncttrue
\mciteSetBstMidEndSepPunct{\mcitedefaultmidpunct}
{\mcitedefaultendpunct}{\mcitedefaultseppunct}\relax
\EndOfBibitem
\bibitem[Woon and Dunning(1995)Woon, and Dunning]{DUNNING_3}
Woon,~D.~E.; Dunning,~T. H.~J. The Pronounced Effect of Microsolvation on
  Diatomic Alkali Halides: Ab Initio Modeling of MX(H2O)n (M = Li, Na; X=F, Cl;
  n = 1-3). \emph{Journal of the American Chemical Society} \textbf{1995},
  \emph{117}, 1090--1097\relax
\mciteBstWouldAddEndPuncttrue
\mciteSetBstMidEndSepPunct{\mcitedefaultmidpunct}
{\mcitedefaultendpunct}{\mcitedefaultseppunct}\relax
\EndOfBibitem
\bibitem[Balakina and Nefediev(2007)Balakina, and Nefediev]{BALAKINA}
Balakina,~M.; Nefediev,~S. The choice of basis set for calculations of linear
  and nonlinear optical properties of conjugated organic molecules in gas and
  in dielectric medium by the example of p-nitroaniline. \emph{Computational
  Materials Science} \textbf{2007}, \emph{38}, 467--472, Selected papers from
  the International Conference on Computational Methods in Sciences and
  Engineering 2004\relax
\mciteBstWouldAddEndPuncttrue
\mciteSetBstMidEndSepPunct{\mcitedefaultmidpunct}
{\mcitedefaultendpunct}{\mcitedefaultseppunct}\relax
\EndOfBibitem
\bibitem[Parsons \latin{et~al.}(2022)Parsons, Balduf, Cheeseman, and
  Caricato]{Parsons}
Parsons,~T.; Balduf,~T.; Cheeseman,~J.~R.; Caricato,~M. Basis Set Dependence of
  Optical Rotation Calculations with Different Choices of Gauge. \emph{The
  Journal of Physical Chemistry A} \textbf{2022}, \emph{126}, 1861--1870, PMID:
  35271772\relax
\mciteBstWouldAddEndPuncttrue
\mciteSetBstMidEndSepPunct{\mcitedefaultmidpunct}
{\mcitedefaultendpunct}{\mcitedefaultseppunct}\relax
\EndOfBibitem
\bibitem[Reis and Papadopoulos(1999)Reis, and Papadopoulos]{reis}
Reis,~H.; Papadopoulos,~M.~G. Nonlinear optical properties of the rhombic
  B4-cluster. \emph{Journal of Computational Chemistry} \textbf{1999},
  \emph{20}, 679--687\relax
\mciteBstWouldAddEndPuncttrue
\mciteSetBstMidEndSepPunct{\mcitedefaultmidpunct}
{\mcitedefaultendpunct}{\mcitedefaultseppunct}\relax
\EndOfBibitem
\bibitem[Lauderdale and Coolidge(1995)Lauderdale, and Coolidge]{Lauderdale}
Lauderdale,~W.~J.; Coolidge,~M.~B. Basis set effects on the nonlinear optical
  properties of selected linear diacetylenes. \emph{The Journal of Physical
  Chemistry} \textbf{1995}, \emph{99}, 9368--9373\relax
\mciteBstWouldAddEndPuncttrue
\mciteSetBstMidEndSepPunct{\mcitedefaultmidpunct}
{\mcitedefaultendpunct}{\mcitedefaultseppunct}\relax
\EndOfBibitem
\bibitem[Jabłoński and Palusiak(2010)Jabłoński, and Palusiak]{Jablonski}
Jabłoński,~M.; Palusiak,~M. Basis Set and Method Dependence in Quantum Theory
  of Atoms in Molecules Calculations for Covalent Bonds. \emph{The Journal of
  Physical Chemistry A} \textbf{2010}, \emph{114}, 12498--12505, PMID:
  21049895\relax
\mciteBstWouldAddEndPuncttrue
\mciteSetBstMidEndSepPunct{\mcitedefaultmidpunct}
{\mcitedefaultendpunct}{\mcitedefaultseppunct}\relax
\EndOfBibitem
\bibitem[Frisch \latin{et~al.}(2016)Frisch, Trucks, Schlegel, Scuseria, Robb,
  Cheeseman, Scalmani, Barone, Petersson, Nakatsuji, Li, Caricato, Marenich,
  Bloino, Janesko, Gomperts, Mennucci, Hratchian, Ortiz, Izmaylov, Sonnenberg,
  Williams-Young, Ding, Lipparini, Egidi, Goings, Peng, Petrone, Henderson,
  Ranasinghe, Zakrzewski, Gao, Rega, Zheng, Liang, Hada, Ehara, Toyota, Fukuda,
  Hasegawa, Ishida, Nakajima, Honda, Kitao, Nakai, Vreven, Throssell,
  Montgomery, Peralta, Ogliaro, Bearpark, Heyd, Brothers, Kudin, Staroverov,
  Keith, Kobayashi, Normand, Raghavachari, Rendell, Burant, Iyengar, Tomasi,
  Cossi, Millam, Klene, Adamo, Cammi, Ochterski, Martin, Morokuma, Farkas,
  Foresman, and Fox]{g16}
Frisch,~M.~J. \latin{et~al.}  Gaussian~16 {R}evision {C}.01. 2016\relax
\mciteBstWouldAddEndPuncttrue
\mciteSetBstMidEndSepPunct{\mcitedefaultmidpunct}
{\mcitedefaultendpunct}{\mcitedefaultseppunct}\relax
\EndOfBibitem
\bibitem[McMurchie \latin{et~al.}()McMurchie, Elbert, Langhoff, and
  Davidson]{meld}
McMurchie,~L.~E.; Elbert,~S.~T.; Langhoff,~S.~R.; Davidson,~E.~R. {MELD}
  package from Indiana University. It has been modified by us to handle bigger
  systems.\relax
\mciteBstWouldAddEndPunctfalse
\mciteSetBstMidEndSepPunct{\mcitedefaultmidpunct}
{}{\mcitedefaultseppunct}\relax
\EndOfBibitem
\bibitem[Shinde and Shukla(2014)Shinde, and Shukla]{C4CP02232G}
Shinde,~R.; Shukla,~A. Large-scale first principles configuration interaction
  calculations of optical absorption in aluminum clusters. \emph{Phys. Chem.
  Chem. Phys.} \textbf{2014}, \emph{16}, 20714--20723\relax
\mciteBstWouldAddEndPuncttrue
\mciteSetBstMidEndSepPunct{\mcitedefaultmidpunct}
{\mcitedefaultendpunct}{\mcitedefaultseppunct}\relax
\EndOfBibitem
\bibitem[Rai \latin{et~al.}(2018)Rai, Chakraborty, and
  Shukla]{doi:10.1021/acs.jpcc.7b08695}
Rai,~D.~K.; Chakraborty,~H.; Shukla,~A. Tunable Optoelectronic Properties of
  Triply Bonded Carbon Molecules with Linear and Graphyne Substructures.
  \emph{The Journal of Physical Chemistry C} \textbf{2018}, \emph{122},
  1309--1317\relax
\mciteBstWouldAddEndPuncttrue
\mciteSetBstMidEndSepPunct{\mcitedefaultmidpunct}
{\mcitedefaultendpunct}{\mcitedefaultseppunct}\relax
\EndOfBibitem
\bibitem[Chakraborty and Shukla(2013)Chakraborty, and
  Shukla]{doi:10.1021/jp408535u}
Chakraborty,~H.; Shukla,~A. Pariser - Parr - Pople Model Based Investigation of
  Ground and Low - Lying Excited States of Long Acenes. \emph{The Journal of
  Physical Chemistry A} \textbf{2013}, \emph{117}, 14220--14229\relax
\mciteBstWouldAddEndPuncttrue
\mciteSetBstMidEndSepPunct{\mcitedefaultmidpunct}
{\mcitedefaultendpunct}{\mcitedefaultseppunct}\relax
\EndOfBibitem
\bibitem[Aryanpour \latin{et~al.}(2014)Aryanpour, Shukla, and
  Mazumdar]{doi:10.1063/1.4867363}
Aryanpour,~K.; Shukla,~A.; Mazumdar,~S. Electron correlations and two-photon
  states in polycyclic aromatic hydrocarbon molecules: A peculiar role of
  geometry. \emph{The Journal of Chemical Physics} \textbf{2014}, \emph{140},
  104301\relax
\mciteBstWouldAddEndPuncttrue
\mciteSetBstMidEndSepPunct{\mcitedefaultmidpunct}
{\mcitedefaultendpunct}{\mcitedefaultseppunct}\relax
\EndOfBibitem
\bibitem[Chakraborty and Shukla(2014)Chakraborty, and
  Shukla]{doi:10.1063/1.4897955}
Chakraborty,~H.; Shukla,~A. Theory of triplet optical absorption in
  oligoacenes: From naphthalene to heptacene. \emph{The Journal of Chemical
  Physics} \textbf{2014}, \emph{141}, 164301\relax
\mciteBstWouldAddEndPuncttrue
\mciteSetBstMidEndSepPunct{\mcitedefaultmidpunct}
{\mcitedefaultendpunct}{\mcitedefaultseppunct}\relax
\EndOfBibitem
\bibitem[SHINDE and SHUKLA(2012)SHINDE, and
  SHUKLA]{doi:10.1142/S1793984411000529}
SHINDE,~R.; SHUKLA,~A. LARGE-SCALE FIRST PRINCIPLES CONFIGURATION INTERACTION
  CALCULATIONS OF OPTICAL ABSORPTION IN BORON CLUSTERS. \emph{Nano LIFE}
  \textbf{2012}, \emph{02}, 1240004\relax
\mciteBstWouldAddEndPuncttrue
\mciteSetBstMidEndSepPunct{\mcitedefaultmidpunct}
{\mcitedefaultendpunct}{\mcitedefaultseppunct}\relax
\EndOfBibitem
\bibitem[Shukla(2002)]{PhysRevB.65.125204}
Shukla,~A. Correlated theory of triplet photoinduced absorption in
  phenylene-vinylene chains. \emph{Phys. Rev. B} \textbf{2002}, \emph{65},
  125204\relax
\mciteBstWouldAddEndPuncttrue
\mciteSetBstMidEndSepPunct{\mcitedefaultmidpunct}
{\mcitedefaultendpunct}{\mcitedefaultseppunct}\relax
\EndOfBibitem
\bibitem[Shukla(2004)]{PhysRevB.69.165218}
Shukla,~A. Theory of nonlinear optical properties of phenyl-substituted
  polyacetylenes. \emph{Phys. Rev. B} \textbf{2004}, \emph{69}, 165218\relax
\mciteBstWouldAddEndPuncttrue
\mciteSetBstMidEndSepPunct{\mcitedefaultmidpunct}
{\mcitedefaultendpunct}{\mcitedefaultseppunct}\relax
\EndOfBibitem
\bibitem[Sony and Shukla(2007)Sony, and Shukla]{PhysRevB.75.155208}
Sony,~P.; Shukla,~A. Large-scale correlated calculations of linear optical
  absorption and low-lying excited states of polyacenes: Pariser-Parr-Pople
  Hamiltonian. \emph{Phys. Rev. B} \textbf{2007}, \emph{75}, 155208\relax
\mciteBstWouldAddEndPuncttrue
\mciteSetBstMidEndSepPunct{\mcitedefaultmidpunct}
{\mcitedefaultendpunct}{\mcitedefaultseppunct}\relax
\EndOfBibitem
\bibitem[Basak \latin{et~al.}(2015)Basak, Chakraborty, and
  Shukla]{PhysRevB.92.205404}
Basak,~T.; Chakraborty,~H.; Shukla,~A. Theory of linear optical absorption in
  diamond-shaped graphene quantum dots. \emph{Phys. Rev. B} \textbf{2015},
  \emph{92}, 205404\relax
\mciteBstWouldAddEndPuncttrue
\mciteSetBstMidEndSepPunct{\mcitedefaultmidpunct}
{\mcitedefaultendpunct}{\mcitedefaultseppunct}\relax
\EndOfBibitem
\bibitem[Priya \latin{et~al.}(2017)Priya, Rai, and Shukla]{Priya2017}
Priya,~P.~K.; Rai,~D.~K.; Shukla,~A. Photoabsorption in sodium clusters: first
  principles configuration interaction calculations. \emph{The European
  Physical Journal D} \textbf{2017}, \emph{71}, 116\relax
\mciteBstWouldAddEndPuncttrue
\mciteSetBstMidEndSepPunct{\mcitedefaultmidpunct}
{\mcitedefaultendpunct}{\mcitedefaultseppunct}\relax
\EndOfBibitem
\bibitem[Shinde and Shukla(2017)Shinde, and Shukla]{Shinde2017}
Shinde,~R.; Shukla,~A. First principles electron-correlated calculations of
  optical absorption in magnesium clusters. \emph{The European Physical Journal
  D} \textbf{2017}, \emph{71}, 301\relax
\mciteBstWouldAddEndPuncttrue
\mciteSetBstMidEndSepPunct{\mcitedefaultmidpunct}
{\mcitedefaultendpunct}{\mcitedefaultseppunct}\relax
\EndOfBibitem
\bibitem[Bhattacharyya \latin{et~al.}(2019)Bhattacharyya, Rai, and
  Shukla]{Bhattacharyya_2019}
Bhattacharyya,~P.; Rai,~D.~K.; Shukla,~A. Systematic First-Principles
  Configuration-Interaction Calculations of Linear Optical Absorption Spectra
  in Silicon Hydrides: Si2H2n (n = 1-3). \emph{The Journal of Physical
  Chemistry A} \textbf{2019}, \emph{123}, 8619--8631\relax
\mciteBstWouldAddEndPuncttrue
\mciteSetBstMidEndSepPunct{\mcitedefaultmidpunct}
{\mcitedefaultendpunct}{\mcitedefaultseppunct}\relax
\EndOfBibitem
\bibitem[Wheeler \latin{et~al.}(2004)Wheeler, Sattelmeyer, Schleyer, and
  Schaefer]{Wheeler_et_al}
Wheeler,~S.~E.; Sattelmeyer,~K.~W.; Schleyer,~P. v.~R.; Schaefer,~H.~F. Binding
  energies of small lithium clusters (Lin) and hydrogenated lithium clusters
  (LinH). \emph{The Journal of Chemical Physics} \textbf{2004}, \emph{120},
  4683--4689\relax
\mciteBstWouldAddEndPuncttrue
\mciteSetBstMidEndSepPunct{\mcitedefaultmidpunct}
{\mcitedefaultendpunct}{\mcitedefaultseppunct}\relax
\EndOfBibitem
\bibitem[Florez and Fuentealba(2009)Florez, and Fuentealba]{Florez_et_al}
Florez,~E.; Fuentealba,~P. A theoretical study of alkali metal atomic clusters:
  From Lin to Csn (n = 2-8). \emph{International Journal of Quantum Chemistry}
  \textbf{2009}, \emph{109}, 1080--1093\relax
\mciteBstWouldAddEndPuncttrue
\mciteSetBstMidEndSepPunct{\mcitedefaultmidpunct}
{\mcitedefaultendpunct}{\mcitedefaultseppunct}\relax
\EndOfBibitem
\bibitem[HUBER(1979)]{Huber}
HUBER,~K.~P. Molecular Structure Constants of Diatomic molecules.
  \emph{Molecular Spectra and molecular Structure Constants of Diatomic
  molecules} \textbf{1979}, \relax
\mciteBstWouldAddEndPunctfalse
\mciteSetBstMidEndSepPunct{\mcitedefaultmidpunct}
{}{\mcitedefaultseppunct}\relax
\EndOfBibitem
\bibitem[Jones \latin{et~al.}(1997)Jones, Lichtenstein, and
  Hutter]{Jones_et_al}
Jones,~R.~O.; Lichtenstein,~A.~I.; Hutter,~J. Density functional study of
  structure and bonding in lithium clusters Lin and their oxides LinO.
  \emph{The Journal of Chemical Physics} \textbf{1997}, \emph{106},
  4566--4574\relax
\mciteBstWouldAddEndPuncttrue
\mciteSetBstMidEndSepPunct{\mcitedefaultmidpunct}
{\mcitedefaultendpunct}{\mcitedefaultseppunct}\relax
\EndOfBibitem
\bibitem[Srinivas and Jellinek(2004)Srinivas, and Jellinek]{Srinivas_et_al}
Srinivas,~S.; Jellinek,~J. Structural and electronic properties of small
  beryllium clusters: A theoretical study. \emph{The Journal of Chemical
  Physics} \textbf{2004}, \emph{121}, 7243--7252\relax
\mciteBstWouldAddEndPuncttrue
\mciteSetBstMidEndSepPunct{\mcitedefaultmidpunct}
{\mcitedefaultendpunct}{\mcitedefaultseppunct}\relax
\EndOfBibitem
\bibitem[Hanley \latin{et~al.}(1988)Hanley, Whitten, and
  Anderson]{Hanley_et_al}
Hanley,~L.; Whitten,~J.~L.; Anderson,~S.~L. Collision-induced dissociation and
  ab initio studies of boron cluster ions: determination of structures and
  stabilities. \emph{The Journal of Physical Chemistry} \textbf{1988},
  \emph{92}, 5803--5812\relax
\mciteBstWouldAddEndPuncttrue
\mciteSetBstMidEndSepPunct{\mcitedefaultmidpunct}
{\mcitedefaultendpunct}{\mcitedefaultseppunct}\relax
\EndOfBibitem
\bibitem[Hong and Wang(2011)Hong, and Wang]{Hong_et_al}
Hong,~X.; Wang,~F. TDDFT calculation for photoabsorption spectra of Lin
  (n=2-11,20) clusters. \emph{Physics Letters A} \textbf{2011}, \emph{375},
  1883 -- 1888\relax
\mciteBstWouldAddEndPuncttrue
\mciteSetBstMidEndSepPunct{\mcitedefaultmidpunct}
{\mcitedefaultendpunct}{\mcitedefaultseppunct}\relax
\EndOfBibitem
\bibitem[Blanc \latin{et~al.}(1991)Blanc, Broyer, Chevaleyre, Dugourd,
  K\"{u}hling, Labastie, Ulbricht, Wolf, and W\"{o}ste]{blanc_et_al_1991}
Blanc,~J.; Broyer,~M.; Chevaleyre,~J.; Dugourd,~P.; K\"{u}hling,~H.;
  Labastie,~P.; Ulbricht,~M.; Wolf,~J.~P.; W\"{o}ste,~L. High resolution
  spectroscopy of small metal clusters. \emph{Zeitschrift f\"{u}r Physik D
  Atoms, Molecules and Clusters} \textbf{1991}, \emph{19}, 7--12\relax
\mciteBstWouldAddEndPuncttrue
\mciteSetBstMidEndSepPunct{\mcitedefaultmidpunct}
{\mcitedefaultendpunct}{\mcitedefaultseppunct}\relax
\EndOfBibitem
\end{mcitethebibliography}

\section*{For Table of Contents Only}

\begin{figure}[H]
	
	\includegraphics[scale=0.5]{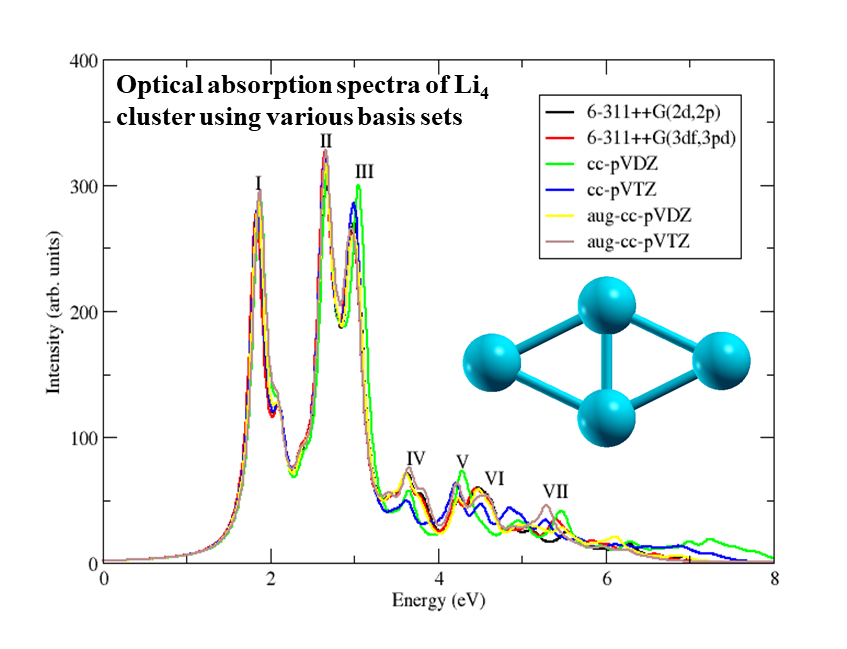}
\end{figure}

\end{document}


\begin{table}
{\tiny{}\caption{Comparison of the peak locations of the optical absorption spectra
of Li$_{2}$ cluster computed using various basis sets. \label{tab:li2-peak-locations}}
}{\tiny\par}

\begin{tabular}{|c|c|c|c|c|c|c|c|c|c|}
\hline 
{\footnotesize{}Basis Set} & {\footnotesize{}Peak I} & {\footnotesize{}Peak II} & {\footnotesize{}Peak III} & {\footnotesize{}Peak IV} & {\footnotesize{}Peak V} & {\footnotesize{}Peak VI} & {\footnotesize{}Peak VII} & {\footnotesize{}Peak VIII} & {\footnotesize{}Peak IX}\tabularnewline
 & {\footnotesize{}(eV)} & {\footnotesize{}(eV)} & {\footnotesize{}(eV)} & {\footnotesize{}(eV)} & {\footnotesize{}(eV)} & {\footnotesize{}(eV)} & {\footnotesize{}(eV)} & {\footnotesize{}(eV)} & {\footnotesize{}(eV)}\tabularnewline
\hline 
{\footnotesize{}6-311++G(2d,2p)} & {\footnotesize{}1.82} & {\footnotesize{}2.61} & {\footnotesize{}3.85} & {\footnotesize{}4.87} & {\footnotesize{}5.88} & {\footnotesize{}6.50} & {\footnotesize{}7.06} & {\footnotesize{}-} & {\footnotesize{}-}\tabularnewline
\hline 
{\footnotesize{}6-311++G(3df,3pd)} & {\footnotesize{}1.83} & {\footnotesize{}2.57} & {\footnotesize{}3.85} & {\footnotesize{}4.61} & {\footnotesize{}4.84} & {\footnotesize{}5.64} & {\footnotesize{}5.88} & {\footnotesize{}6.08} & {\footnotesize{}7.00}\tabularnewline
\hline 
{\footnotesize{}cc-pVDZ} & {\footnotesize{}1.82} & {\footnotesize{}2.65} & {\footnotesize{}4.43} & {\footnotesize{}5.95} & {\footnotesize{}7.12} & {\footnotesize{}-} & {\footnotesize{}-} & {\footnotesize{}-} & {\footnotesize{}-}\tabularnewline
\hline 
{\footnotesize{}cc-pVTZ} & {\footnotesize{}1.83} & {\footnotesize{}2.58} & {\footnotesize{}4.26} & {\footnotesize{}5.43} & {\footnotesize{}5.98} & {\footnotesize{}6.64} & {\footnotesize{}7.03} & {\footnotesize{}-} & {\footnotesize{}-}\tabularnewline
\hline 
{\footnotesize{}aug-cc-pVDZ} & {\footnotesize{}1.82} & {\footnotesize{}2.60} & {\footnotesize{}3.84} & {\footnotesize{}4.58} & {\footnotesize{}5.06} & {\footnotesize{}5.94} & {\footnotesize{}6.67} & {\footnotesize{}6.96} & {\footnotesize{}-}\tabularnewline
\hline 
{\footnotesize{}aug-cc-pVTZ} & {\footnotesize{}1.83} & {\footnotesize{}2.57} & {\footnotesize{}3.87} & {\footnotesize{}4.59} & {\footnotesize{}5.43} & {\footnotesize{}5.93} & {\footnotesize{}6.66} & {\footnotesize{}7.03} & {\footnotesize{}-}\tabularnewline
\hline 
\end{tabular}
\end{table}

\begin{table}
\caption{Comparison of the peak locations of the optical absorption spectra
of Li$_{3}$ chain computed using various basis sets. \label{tab:li3-chain-peak-locations}}

\begin{tabular}{|c|c|c|c|c|c|c|}
\hline 
Basis Set & Peak I & Peak II & Peak III & Peak IV & Peak V & Peak VI\tabularnewline
 & (eV) & (eV) & (eV) & (eV) & (eV) & (eV)\tabularnewline
\hline 
6-311++G(2d,2p) & 0.72 & 1.27 & 2.58 & 3.39 & 3.91 & 4.28\tabularnewline
\hline 
6-311++G(3df,3pd) & 0.72 & 1.27 & 2.54 & 3.38 & 3.90 & 4.14\tabularnewline
\hline 
cc-pVDZ & 0.72 & 1.26 & 2.65 & 3.47 & 4.37 & 4.92\tabularnewline
\hline 
cc-pVTZ & 0.72 & 1.28 & 2.57 & 3.45 & 4.39 & 4.58\tabularnewline
\hline 
aug-cc-pVDZ & 0.72 & 1.26 & 2.56 & 3.38 & 3.88 & -\tabularnewline
\hline 
aug-cc-pVTZ & 0.72 & 1.27 & 2.53 & 3.37 & 3.89 & -\tabularnewline
\hline 
\end{tabular}
\end{table}

\begin{table}
\caption{The peak locations of the optical absorption spectra of Li$_{3}$
isosceles triangular cluster computed using various basis sets are
compared. \label{tab-li3-traingular_peak-locations}}

{\scriptsize{}}%
\begin{tabular}{|c|c|c|c|c|c|c|c|}
\hline 
Basis Set & Peak I & Peak II & Peak III & Peak IV & Peak V & Peak VI & Peak VII\tabularnewline
 & (eV) & (eV) & (eV) & (eV) & (eV) & (eV) & (eV)\tabularnewline
\hline 
6-311++G(2d,2p) & 1.09 & 1.41 & 2.14 & 2.41 & 2.66 & 2.97 & 3.21\tabularnewline
\hline 
6-311++G(3df,3pd) & 1.07 & 1.42 & 2.12 & 2.39 & 2.65 & 2.96 & 3.19\tabularnewline
\hline 
cc-pVDZ & 1.11 & 1.40 & 2.18 & 2.43 & 2.61 & 2.81 & 3.13\tabularnewline
\hline 
cc-pVTZ & 1.08 & 1.42 & 2.15 & 2.40 & 2.70 & 3.02 & 4.27\tabularnewline
\hline 
aug-cc-pVDZ & 1.09 & 1.41 & 2.13 & 2.40 & 2.65 & 2.96 & 3.20\tabularnewline
\hline 
aug-cc-pVTZ & 1.07 & 1.42 & 2.11 & 2.43 & 2.65 & 2.95 & 3.20\tabularnewline
\hline 
\end{tabular}{\scriptsize\par}
\end{table}

\begin{table}
\caption{High energy peak locations of the optical absorption spectra of Li$_{3}$
Isosceles triangular cluster computed using various basis sets are
compared. \label{tab-li3-triangular-high-energy-peaks}}

\begin{tabular}{|c|c|c|c|c|c|}
\hline 
Li$_{3}$ Cluster & Peak VIII & Peak IX & Peak X & Peak XI & Peak XII\tabularnewline
Basis Set & (eV) & (eV) & (eV) & (eV) & (eV)\tabularnewline
\hline 
6-311++G(2d,2p) & 3.77	 & 4.25 & 5.29 & 5.99 & -\tabularnewline
\hline 
6-311++G(3df,3pd) & 3.75	 & 4.17 & 4.91 & 5.25 & 5.79\tabularnewline
\hline 
cc-pVDZ & 3.57	 & 4.24 & 4.94 & 5.33 & 	6.05\tabularnewline
\hline 
cc-pVTZ & 4.69	 & 5.64 & - & - & -\tabularnewline
\hline 
aug-cc-pVDZ & 4.16	 & 4.44 & 5.35 & 5.59 & 5.94\tabularnewline
\hline 
aug-cc-pVTZ & 3.77	 & 4.11 & 5.39 & 5.60 & -\tabularnewline
\hline 
\end{tabular}
\end{table}

\begin{table}
\caption{The peak locations of the optical absorption spectra of Li$_{4}$
rhombus cluster computed using various basis sets. \label{tab:li4-peak-locations}
The higher energy peak locations are presented in the Table I of the
Supporting Information.}

\begin{tabular}{|c|c|c|c|c|c|c|c|}
\hline 
Basis Set & Peak I & Peak II & Peak III & Peak IV & Peak V & Peak VI & Peak VII\tabularnewline
 & (eV) & (eV) & (eV) & (eV) & (eV) & (eV) & (eV)\tabularnewline
\hline 
6-311++G(2d,2p) & 1.84 & 2.65 & 3.05 & 3.64 & 4.24 & 4.52 & 5.15\tabularnewline
\hline 
6-311++G(3df,3pd) & 1.83 & 2.64 & 2.93 & 3.63 & 4.21 & 4.51 & 5.11\tabularnewline
\hline 
cc-pVDZ & 1.87 & 2.67 & 3.08 & 3.65 & 4.27 & 4.86 & 5.42\tabularnewline
\hline 
cc-pVTZ & 1.84 & 2.65 & 3.03 & 3.64 & 4.22 & 4.54 & 5.30\tabularnewline
\hline 
aug-cc-pVDZ & 1.85 & 2.65 & 3.05 & 3.61 & 4.23 & 4.62 & -\tabularnewline
\hline 
aug-cc-pVTZ & 1.87 & 2.65 & 2.93 & 3.66 & 4.22 & 4.61 & 5.30\tabularnewline
\hline 
\end{tabular}
\end{table}

\begin{table}
\caption{The peak locations of the optical absorption spectra of Be$_{2}^{+}$
cluster computed using various basis sets are compared. \label{tab:be2+-peak-locations}}

\begin{tabular}{|c|c|c|c|c|c|c|c|c|c|}
\hline 
{\footnotesize{}Basis Set} & {\footnotesize{}Peak I} & {\footnotesize{}Peak II} & {\footnotesize{}Peak III} & {\footnotesize{}Peak IV} & {\footnotesize{}Peak V} & {\footnotesize{}Peak VI} & {\footnotesize{}Peak VII} & {\footnotesize{}Peak VIII} & {\footnotesize{}Peak IX}\tabularnewline
 & {\footnotesize{}(eV)} & {\footnotesize{}(eV)} & {\footnotesize{}(eV)} & {\footnotesize{}(eV)} & {\footnotesize{}(eV)} & {\footnotesize{}(eV)} & {\footnotesize{}(eV)} & {\footnotesize{}(eV)} & {\footnotesize{}(eV)}\tabularnewline
\hline 
{\footnotesize{}6-311++G(2d,2p)} & {\footnotesize{}1.75} & {\footnotesize{}3.66} & {\footnotesize{}4.18} & {\footnotesize{}6.04} & {\footnotesize{}6.32} & {\footnotesize{}8.32} & {\footnotesize{}8.90} & {\footnotesize{}9.33} & {\footnotesize{}9.65}\tabularnewline
\hline 
{\footnotesize{}6-311++G(3df,3pd)} & {\footnotesize{}1.74} & {\footnotesize{}3.68} & {\footnotesize{}4.18} & {\footnotesize{}6.04} & {\footnotesize{}6.30} & {\footnotesize{}8.33} & {\footnotesize{}8.84} & {\footnotesize{}9.31} & {\footnotesize{}9.63}\tabularnewline
\hline 
{\footnotesize{}cc-pVDZ} & {\footnotesize{}1.76} & {\footnotesize{}3.69} & {\footnotesize{}4.20} & {\footnotesize{}6.12} & {\footnotesize{}6.37} & {\footnotesize{}8.28} & {\footnotesize{}9.61} & {\footnotesize{}-} & {\footnotesize{}-}\tabularnewline
\hline 
{\footnotesize{}cc-pVTZ} & {\footnotesize{}1.74} & {\footnotesize{}3.68} & {\footnotesize{}4.19} & {\footnotesize{}6.05} & {\footnotesize{}6.31} & {\footnotesize{}8.44} & {\footnotesize{}9.50} & {\footnotesize{}-} & {\footnotesize{}-}\tabularnewline
\hline 
{\footnotesize{}aug-cc-pVDZ} & {\footnotesize{}1.77} & {\footnotesize{}3.69} & {\footnotesize{}4.18} & {\footnotesize{}6.08} & {\footnotesize{}6.36} & {\footnotesize{}8.38} & {\footnotesize{}9.09} & {\footnotesize{}9.47} & {\footnotesize{}-}\tabularnewline
\hline 
{\footnotesize{}aug-cc-pVTZ} & {\footnotesize{}1.74} & {\footnotesize{}3.68} & {\footnotesize{}4.18} & {\footnotesize{}6.03} & {\footnotesize{}6.30} & {\footnotesize{}8.34} & {\footnotesize{}8.92} & {\footnotesize{}9.38} & {\footnotesize{}-}\tabularnewline
\hline 
\end{tabular}
\end{table}

\begin{table}
\caption{The peak locations of the optical absorption spectra of Be$_{3}^{+}$
cluster computed using various basis sets are compared. \label{tab:be3+-peak-locations}}

\begin{tabular}{|c|c|c|c|c|c|c|c|}
\hline 
Basis Set & Peak I & Peak II & Peak III & Peak IV & Peak V & Peak VI & Peak VII\tabularnewline
 & (eV) & (eV) & (eV) & (eV) & (eV) & (eV) & (eV)\tabularnewline
\hline 
6-311++G(2d,2p) & 1.05 & 3.16 & 3.66 & 4.90 & 5.39 & 5.86 & 6.53\tabularnewline
\hline 
6-311++G(3df,3pd) & 1.02 & 3.17 & 3.63 & 4.84 & 5.40 & 5.83 & 6.47\tabularnewline
\hline 
cc-pVDZ & 1.04 & 3.16 & 3.67 & 4.91 & 5.39 & 5.89 & 6.52\tabularnewline
\hline 
cc-pVTZ & 1.01 & 3.14 & 3.60 & 4.87 & 5.37 & 5.83 & 6.50\tabularnewline
\hline 
aug-cc-pVDZ & 1.02 & 3.16 & 3.66 & 4.87 & 5.41 & 5.87 & 6.51\tabularnewline
\hline 
aug-cc-pVTZ & 1.02 & 3.16 & 3.66 & 4.87 & 5.40 & 5.85 & 6.52\tabularnewline
\hline 
\end{tabular}
\end{table}

\begin{table}[h]
\caption{The peak locations of the optical absorption spectra of B$_{2}^{+}$
cluster computed using various basis sets are compared. \label{tab:b2+-peak-locations}}

\begin{tabular}{|c|c|c|c|c|c|c|c|c|c|}
\hline 
{\footnotesize{}Basis Set} & {\footnotesize{}Peak I} & {\footnotesize{}Peak II} & {\footnotesize{}Peak III} & {\footnotesize{}Peak IV} & {\footnotesize{}Peak V} & {\footnotesize{}Peak VI} & {\footnotesize{}Peak VII} & {\footnotesize{}Peak VIII} & {\footnotesize{}Peak IX}\tabularnewline
 & {\footnotesize{}(eV)} & {\footnotesize{}(eV)} & {\footnotesize{}(eV)} & {\footnotesize{}(eV)} & {\footnotesize{}(eV)} & {\footnotesize{}(eV)} & {\footnotesize{}(eV)} & {\footnotesize{}(eV)} & {\footnotesize{}(eV)}\tabularnewline
\hline 
{\footnotesize{}6-311++G(2d,2p)} & {\footnotesize{}3.65} & {\footnotesize{}4.86} & {\footnotesize{}6.01} & {\footnotesize{}7.00} & {\footnotesize{}7.63} & {\footnotesize{}9.68} & {\footnotesize{}10.26} & {\footnotesize{}11.42} & {\footnotesize{}12.71}\tabularnewline
\hline 
{\footnotesize{}6-311++G(3df,3pd)} & {\footnotesize{}3.65} & {\footnotesize{}4.77} & {\footnotesize{}5.98} & {\footnotesize{}7.14} & {\footnotesize{}7.62} & {\footnotesize{}9.65} & {\footnotesize{}10.22} & {\footnotesize{}11.40} & {\footnotesize{}12.61}\tabularnewline
\hline 
{\footnotesize{}cc-pVDZ} & {\footnotesize{}3.64} & {\footnotesize{}4.79} & {\footnotesize{}6.03} & {\footnotesize{}7.16} & {\footnotesize{}-} & {\footnotesize{}9.88} & {\footnotesize{}11.29} & {\footnotesize{}12.74} & \tabularnewline
\hline 
{\footnotesize{}cc-pVTZ} & {\footnotesize{}3.66} & {\footnotesize{}4.80} & {\footnotesize{}5.99} & {\footnotesize{}7.05} & {\footnotesize{}7.63} & {\footnotesize{}9.87} & {\footnotesize{}11.12} & {\footnotesize{}12.77} & {\footnotesize{}-}\tabularnewline
\hline 
{\footnotesize{}aug-cc-pVDZ} & {\footnotesize{}3.63} & {\footnotesize{}4.78} & {\footnotesize{}5.98} & {\footnotesize{}7.03} & {\footnotesize{}7.60} & {\footnotesize{}9.66} & {\footnotesize{}10.27} & {\footnotesize{}11.31} & {\footnotesize{}12.52}\tabularnewline
\hline 
{\footnotesize{}aug-cc-pVTZ} & {\footnotesize{}3.66} & {\footnotesize{}4.80} & {\footnotesize{}5.99} & {\footnotesize{}7.04} & {\footnotesize{}7.65} & {\footnotesize{}9.65} & {\footnotesize{}10.21} & {\footnotesize{}11.41} & {\footnotesize{}12.20}\tabularnewline
\hline 
\end{tabular}
\end{table}

\begin{table}
\caption{The peak locations of the optical absorption spectra of B$_{3}^{+}$
cluster computed using various basis sets are compared. \label{tab:b3+-peak-locations}}

\begin{tabular}{|c|c|c|c|c|c|c|c|c|}
\hline 
{\footnotesize{}B$_{3}^{+}$ Cluster} & {\footnotesize{}Peak I} & {\footnotesize{}Peak II} & {\footnotesize{}Peak III} & {\footnotesize{}Peak IV} & {\footnotesize{}Peak V} & {\footnotesize{}Peak VI} & {\footnotesize{}Peak VII} & {\footnotesize{}Peak VIII}\tabularnewline
{\footnotesize{}Basis Set} & {\footnotesize{}(eV)} & {\footnotesize{}(eV)} & {\footnotesize{}(eV)} & {\footnotesize{}(eV)} & {\footnotesize{}(eV)} & {\footnotesize{}(eV)} & {\footnotesize{}(eV)} & {\footnotesize{}(eV)}\tabularnewline
\hline 
{\footnotesize{}6-311++G(2d,2p)} & {\footnotesize{}0.82} & {\footnotesize{}3.24} & {\footnotesize{}5.00} & {\footnotesize{}5.38} & {\footnotesize{}6.02} & {\footnotesize{}7.12} & {\footnotesize{}8.09} & {\footnotesize{}8.85}\tabularnewline
\hline 
{\footnotesize{}6-311++G(3df,3pd)} & {\footnotesize{}0.83} & {\footnotesize{}3.21} & {\footnotesize{}4.99} & {\footnotesize{}5.44} & {\footnotesize{}6.06} & {\footnotesize{}7.17} & {\footnotesize{}8.07} & {\footnotesize{}8.87}\tabularnewline
\hline 
{\footnotesize{}cc-pVDZ} & {\footnotesize{}0.96} & {\footnotesize{}3.28} & {\footnotesize{}5.17} & {\footnotesize{}5.59} & {\footnotesize{}6.11} & {\footnotesize{}7.39} & {\footnotesize{}8.21} & {\footnotesize{}8.95}\tabularnewline
\hline 
{\footnotesize{}cc-pVTZ} & {\footnotesize{}0.82} & {\footnotesize{}3.23} & {\footnotesize{}4.98} & {\footnotesize{}5.35} & {\footnotesize{}6.03} & {\footnotesize{}7.01} & {\footnotesize{}8.06} & {\footnotesize{}8.79}\tabularnewline
\hline 
{\footnotesize{}aug-cc-pVDZ} & {\footnotesize{}0.84} & {\footnotesize{}3.22} & {\footnotesize{}5.00} & {\footnotesize{}5.41} & {\footnotesize{}6.02} & {\footnotesize{}7.10} & {\footnotesize{}8.03} & {\footnotesize{}8.80}\tabularnewline
\hline 
{\footnotesize{}aug-cc-pVTZ} & {\footnotesize{}0.84} & {\footnotesize{}3.22} & {\footnotesize{}5.01} & {\footnotesize{}5.45} & {\footnotesize{}6.03} & {\footnotesize{}7.16} & {\footnotesize{}8.08} & {\footnotesize{}8.85}\tabularnewline
\hline 
\end{tabular}
\end{table}

\begin{table}[H]
\caption{The peak locations of the optical absorption spectra of B$_{3}^{+}$
cluster employing TD-DFT method with B3LYP functional and computed
using various basis sets are compared. \label{tab:DFT_b3+-peak-locations}}

\begin{tabular}{|c|c|c|c|c|}
\hline 
{\footnotesize{}B$_{3}^{+}$ Cluster} & {\footnotesize{}Peak I} & {\footnotesize{}Peak II} & {\footnotesize{}Peak III} & {\footnotesize{}Peak IV}\tabularnewline
{\footnotesize{}Basis Set} & {\footnotesize{}(eV)} & {\footnotesize{}(eV)} & {\footnotesize{}(eV)} & {\footnotesize{}(eV)}\tabularnewline
\hline 
{\footnotesize{}6-311++G(2d,2p)} & 0.94  & 2.95  & 5.70  & 9.35 \tabularnewline
\hline 
{\footnotesize{}6-311++G(3df,3pd)} & 0.95 & 2.94 & 5.70 & 9.35\tabularnewline
\hline 
{\footnotesize{}cc-pVDZ} & 0.96 & 3.00 & 5.73 & 9.52\tabularnewline
\hline 
{\footnotesize{}cc-pVTZ} & 0.96 & 2.95 & 5.70 & 9.40\tabularnewline
\hline 
{\footnotesize{}aug-cc-pVDZ} & 0.96 & 2.99 & 5.73 & 9.40\tabularnewline
\hline 
{\footnotesize{}aug-cc-pVTZ} & 0.95 & 2.94 & 5.70 & 9.36\tabularnewline
\hline 
\end{tabular}
\end{table}

\begin{table}[H]
\caption{Comparison of oscillator strengths and the dominant configurations
contributing to the many-particle wave functions for peaks I and the
maximum intensity peak (peak III) of Li$_{3}$ chain calculated using
different basis sets. In the “Polarization” column, $\parallel$ indicates
photon polarization along the direction of the molecule (longitudinal
polarization), while $\perp$ indicates polarization perpendicular
to the molecular axis (transverse polarization). ’H’ and ’L’ stand
for HOMO and LUMO orbitals. \label{tab:osc-str-li3-linear}}

\centering{}{\small{}}%
\begin{tabular}{|c|c|c|c|c|c|c|}
\hline 
{\scriptsize{}Basis Set} & \multicolumn{1}{c}{} & \multicolumn{1}{c}{{\scriptsize{}Peak I}} &  & \multicolumn{1}{c}{} & \multicolumn{1}{c}{{\scriptsize{}Peak III}} & \tabularnewline
\cline{2-7} \cline{3-7} \cline{4-7} \cline{5-7} \cline{6-7} \cline{7-7} 
 & {\scriptsize{}Polarization} & {\scriptsize{}$f$} & {\scriptsize{}Wave-function} & {\scriptsize{}Polarization} & {\scriptsize{}$f$} & {\scriptsize{}Wave-function}\tabularnewline
\hline 
\hline 
{\scriptsize{}6-311++G(2d,2p)} & {\scriptsize{}$\parallel$} & {\scriptsize{}0.106} & {\scriptsize{}$\arrowvert H-1\rightarrow H\rangle$} & {\scriptsize{}$\perp$} & {\scriptsize{}0.591} & {\scriptsize{}$\arrowvert H-1\rightarrow L+2\rangle$}\tabularnewline
 &  &  & {\scriptsize{}$\arrowvert H\rightarrow L+6\rangle$} &  &  & \tabularnewline
\hline 
{\scriptsize{}6-311++G(3df,3pd)} & {\scriptsize{}$\parallel$} & {\scriptsize{}0.106} & {\scriptsize{}$\arrowvert H-1\rightarrow H\rangle$} & {\scriptsize{}$\perp$} & {\scriptsize{}0.595} & {\scriptsize{}$\arrowvert H-1\rightarrow L+2\rangle$}\tabularnewline
 &  &  & {\scriptsize{}$\arrowvert H\rightarrow L+5\rangle$} &  &  & {\scriptsize{}$\arrowvert H\rightarrow L+11\rangle$}\tabularnewline
\hline 
{\scriptsize{}cc-PVDZ} & {\scriptsize{}$\parallel$} & {\scriptsize{}0.097} & {\scriptsize{}$\arrowvert H-1\rightarrow H\rangle$} & {\scriptsize{}$\perp$} & {\scriptsize{}0.631} & {\scriptsize{}$\arrowvert H-1\rightarrow L+1\rangle$}\tabularnewline
 &  &  & {\scriptsize{}$\arrowvert H\rightarrow L\rangle$} &  &  & {\scriptsize{}$\arrowvert H\rightarrow L+3\rangle$}\tabularnewline
\hline 
{\scriptsize{}cc-pVTZ} & {\scriptsize{}$\parallel$} & {\scriptsize{}0.105} & {\scriptsize{}$\arrowvert H-1\rightarrow H\rangle$} & {\scriptsize{}$\perp$} & {\scriptsize{}0.618} & {\scriptsize{}$\arrowvert H-1\rightarrow L+1\rangle$}\tabularnewline
 &  &  & {\scriptsize{}$\arrowvert H\rightarrow L\rangle$} &  &  & {\scriptsize{}$\arrowvert H\rightarrow L+3\rangle$}\tabularnewline
\hline 
{\scriptsize{}aug-cc-pVDZ} & {\scriptsize{}$\parallel$} & {\scriptsize{}0.101} & {\scriptsize{}$\arrowvert H-1\rightarrow H\rangle$} & {\scriptsize{}$\perp$} & {\scriptsize{}0.603} & {\scriptsize{}$\arrowvert H-1\rightarrow L+2\rangle$}\tabularnewline
 &  &  & {\scriptsize{}$\arrowvert H\rightarrow L+7\rangle$} &  &  & {\scriptsize{}$\arrowvert H\rightarrow L+10\rangle$}\tabularnewline
\hline 
{\scriptsize{}aug-cc-pVTZ} & {\scriptsize{}$\parallel$} & {\scriptsize{}0.106} & {\scriptsize{}$\arrowvert HF\rangle$} & {\scriptsize{}$\perp$} & {\scriptsize{}0.594} & {\scriptsize{}$\arrowvert H-1\rightarrow L+2\rangle$}\tabularnewline
 &  &  & {\scriptsize{}$\arrowvert H-1\rightarrow H;$} &  &  & {\scriptsize{}$\arrowvert H-1\rightarrow H;$}\tabularnewline
 &  &  & {\scriptsize{}$H-1\rightarrow L+5\rangle$} &  &  & {\scriptsize{}$H-1\rightarrow L+11\rangle$}\tabularnewline
\hline 
\end{tabular}{\small\par}
\end{table}

\begin{table}[H]
\caption{Comparison of oscillator strengths and the dominant configurations
contributing to the many-particle wave functions for peaks I and the
maximum intensity peak (peak IV) of Li$_{3}$ triangular calculated
using different basis sets. The rest of the information is same as
in the caption of Table \ref{tab:osc-str-li3-linear}. \label{tab:Osc-str-li3-triangular}}

\centering{}%
\begin{tabular}{|c|c|c|c|c|c|c|}
\hline 
{\scriptsize{}Basis Set} & \multicolumn{1}{c}{} & \multicolumn{1}{c}{{\scriptsize{}Peak I}} &  & \multicolumn{1}{c}{} & \multicolumn{1}{c}{{\scriptsize{}Peak IV}} & \tabularnewline
\cline{2-7} \cline{3-7} \cline{4-7} \cline{5-7} \cline{6-7} \cline{7-7} 
 & {\scriptsize{}Polarization} & {\scriptsize{}$f$} & {\scriptsize{}Wave-function} & {\scriptsize{}Polarization} & {\scriptsize{}$f$} & {\scriptsize{}Wave-function}\tabularnewline
\hline 
\hline 
{\scriptsize{}6-311++G(2d,2p)} & {\scriptsize{}$\parallel$} & {\scriptsize{}0.127} & {\scriptsize{}$\arrowvert H\rightarrow L+14\rangle$} & {\scriptsize{}$\parallel$} & {\scriptsize{}0.465} & {\scriptsize{}$\arrowvert H-1\rightarrow L\rangle$}\tabularnewline
 &  &  & {\scriptsize{}$\arrowvert H\rightarrow L+1\rangle$} &  &  & {\scriptsize{}$\arrowvert H\rightarrow L+9\rangle$}\tabularnewline
\hline 
{\scriptsize{}6-311++G(3df,3pd)} & {\scriptsize{}$\parallel$} & {\scriptsize{}0.131} & {\scriptsize{}$\arrowvert H\rightarrow L+14\rangle$} & {\scriptsize{}$\parallel$} & {\scriptsize{}0.459} & {\scriptsize{}$\arrowvert H-1\rightarrow L\rangle$}\tabularnewline
 &  &  & {\scriptsize{}$\arrowvert H\rightarrow L+1\rangle$} &  &  & {\scriptsize{}$\arrowvert H\rightarrow L+5\rangle$}\tabularnewline
\hline 
{\scriptsize{}cc-PVDZ} & {\scriptsize{}$\parallel$} & {\scriptsize{}0.121} & {\scriptsize{}$\arrowvert H\rightarrow L+2\rangle$} & {\scriptsize{}$\parallel$} & {\scriptsize{}0.488} & {\scriptsize{}$\arrowvert H-1\rightarrow L\rangle$}\tabularnewline
 &  &  & {\scriptsize{}$\arrowvert H-1\rightarrow L+2\rangle$} &  &  & {\scriptsize{}$\arrowvert H\rightarrow L+4\rangle$}\tabularnewline
\hline 
{\scriptsize{}cc-pVTZ} & {\scriptsize{}$\parallel$} & {\scriptsize{}0.129} & {\scriptsize{}$\arrowvert H\rightarrow L+2\rangle$} & {\scriptsize{}$\parallel$} & {\scriptsize{}0.478} & {\scriptsize{}$\arrowvert H-1\rightarrow L\rangle$}\tabularnewline
 &  &  & {\scriptsize{}$\arrowvert H-1\rightarrow L+2\rangle$} &  &  & {\scriptsize{}$\arrowvert H\rightarrow L+4\rangle$}\tabularnewline
\hline 
{\scriptsize{}aug-cc-pVDZ} & {\scriptsize{}$\parallel$} & {\scriptsize{}0.126} & {\scriptsize{}$\arrowvert H\rightarrow L+14\rangle$} & {\scriptsize{}$\parallel$} & {\scriptsize{}0.469} & {\scriptsize{}$\arrowvert H-1\rightarrow L\rangle$}\tabularnewline
 &  &  & {\scriptsize{}$\arrowvert H\rightarrow L+10\rangle$} &  &  & {\scriptsize{}$\arrowvert H\rightarrow L+16\rangle$}\tabularnewline
\hline 
{\scriptsize{}aug-cc-pVTZ} & {\scriptsize{}$\parallel$} & {\scriptsize{}0.129} & {\scriptsize{}$\arrowvert H\rightarrow L+13\rangle$} & {\scriptsize{}$\parallel$} & {\scriptsize{}0.462} & {\scriptsize{}$\arrowvert H-1\rightarrow L\rangle$}\tabularnewline
 &  &  & {\scriptsize{}$\arrowvert H\rightarrow L+1\rangle$} &  &  & {\scriptsize{}$\arrowvert H\rightarrow L+5\rangle$}\tabularnewline
\hline 
\end{tabular}
\end{table}

\begin{table}[H]
\caption{Comparison of oscillator strengths and the dominant configurations
contributing to the many-particle wave functions for peaks I and the
maximum intensity peak (peak II) of Li$_{4}$ cluster calculated using
different basis sets. The rest of the information is same as in the
caption of Table \ref{tab:osc-str-li3-linear}.\label{tab:Osc-strength-li4}}

\centering{}%
\begin{tabular}{|c|c|c|c|c|c|c|}
\hline 
{\scriptsize{}Basis Set} & \multicolumn{1}{c}{} & \multicolumn{1}{c}{{\scriptsize{}Peak I}} &  & \multicolumn{1}{c}{} & \multicolumn{1}{c}{{\scriptsize{}Peak II}} & \tabularnewline
\cline{2-7} \cline{3-7} \cline{4-7} \cline{5-7} \cline{6-7} \cline{7-7} 
 & {\scriptsize{}Polarization} & {\scriptsize{}$f$} & {\scriptsize{}Wave-function} & {\scriptsize{}Polarization} & {\scriptsize{}$f$} & {\scriptsize{}Wave-function}\tabularnewline
\hline 
\hline 
{\scriptsize{}6-311++G(2d,2p)} & {\scriptsize{}$\parallel$} & {\scriptsize{}0.628} & {\scriptsize{}$\arrowvert H\rightarrow L+1\rangle$} & {\scriptsize{}$\parallel$} & {\scriptsize{}0.648} & {\scriptsize{}$\arrowvert H-1\rightarrow L\rangle$}\tabularnewline
 &  &  & {\scriptsize{}$\arrowvert H\rightarrow L+8\rangle$} &  &  & {\scriptsize{}$\arrowvert H-1\rightarrow L+5\rangle$}\tabularnewline
\hline 
{\scriptsize{}6-311++G(3df,3pd)} & {\scriptsize{}$\parallel$} & {\scriptsize{}0.615} & {\scriptsize{}$\arrowvert H\rightarrow L+1\rangle$} & {\scriptsize{}$\parallel$} & {\scriptsize{}0.683} & {\scriptsize{}$\arrowvert H-1\rightarrow L\rangle$}\tabularnewline
 &  &  & {\scriptsize{}$\arrowvert H\rightarrow L+9\rangle$} &  &  & {\scriptsize{}$\arrowvert H-1\rightarrow L+6\rangle$}\tabularnewline
\hline 
{\scriptsize{}cc-PVDZ} & {\scriptsize{}$\parallel$} & {\scriptsize{}0.659} & {\scriptsize{}$\arrowvert H\rightarrow L+1\rangle$} & {\scriptsize{}$\parallel$} & {\scriptsize{}0.649} & {\scriptsize{}$\arrowvert H-1\rightarrow L\rangle$}\tabularnewline
 &  &  & {\scriptsize{}$\arrowvert H\rightarrow L+4\rangle$} &  &  & {\scriptsize{}$\arrowvert H\rightarrow L+3\rangle$}\tabularnewline
\hline 
{\scriptsize{}cc-pVTZ} & {\scriptsize{}$\parallel$} & {\scriptsize{}0.624} & {\scriptsize{}$\arrowvert H\rightarrow L+1\rangle$} & {\scriptsize{}$\parallel$} & {\scriptsize{}0.678} & {\scriptsize{}$\arrowvert H-1\rightarrow L\rangle$}\tabularnewline
 &  &  & {\scriptsize{}$\arrowvert H\rightarrow L+5\rangle$} &  &  & {\scriptsize{}$\arrowvert H\rightarrow L+4\rangle$}\tabularnewline
\hline 
{\scriptsize{}aug-cc-pVDZ} & {\scriptsize{}$\parallel$} & {\scriptsize{}0.634} & {\scriptsize{}$\arrowvert H\rightarrow L+1\rangle$} & {\scriptsize{}$\parallel$} & {\scriptsize{}0.654} & {\scriptsize{}$\arrowvert H-1\rightarrow L\rangle$}\tabularnewline
 &  &  & {\scriptsize{}$\arrowvert H\rightarrow L+9\rangle$} &  &  & {\scriptsize{}$\arrowvert H-1\rightarrow L+5\rangle$}\tabularnewline
\hline 
{\scriptsize{}aug-cc-pVTZ} & {\scriptsize{}$\parallel$} & {\scriptsize{}0.656} & {\scriptsize{}$\arrowvert H\rightarrow L+1\rangle$} & {\scriptsize{}$\parallel$} & {\scriptsize{}0.660} & {\scriptsize{}$\arrowvert H-1\rightarrow L\rangle$}\tabularnewline
 &  &  & {\scriptsize{}$\arrowvert H\rightarrow L+9\rangle$} &  &  & {\scriptsize{}$\arrowvert H-1\rightarrow L+5\rangle$}\tabularnewline
\hline 
\end{tabular}
\end{table}

\begin{table}[H]
\caption{Comparison of oscillator strengths and the dominant configurations
contributing to the many-particle wave functions for peaks I and the
maximum intensity peak (peak V) of Be$_{2}^{+}$ cluster calculated
using different basis sets. The rest of the information is same as
in the caption of Table \ref{tab:osc-str-li3-linear}\label{tab:Osc-strength-be2+}}

\centering{}%
\begin{tabular}{|c|c|c|c|c|c|c|}
\hline 
{\scriptsize{}Basis Set} & \multicolumn{1}{c}{} & \multicolumn{1}{c}{{\scriptsize{}Peak I}} &  & \multicolumn{1}{c}{} & \multicolumn{1}{c}{{\scriptsize{}Peak V}} & \tabularnewline
\cline{2-7} \cline{3-7} \cline{4-7} \cline{5-7} \cline{6-7} \cline{7-7} 
 & {\scriptsize{}Polarization} & {\scriptsize{}$f$} & {\scriptsize{}Wave-function} & {\scriptsize{}Polarization} & {\scriptsize{}$f$} & {\scriptsize{}Wave-function}\tabularnewline
\hline 
\hline 
{\scriptsize{}6-311++G(2d,2p)} & {\scriptsize{}$\parallel$} & {\scriptsize{}0.113} & {\scriptsize{}$\arrowvert H\rightarrow L\rangle$} & {\scriptsize{}$\perp$} & {\scriptsize{}0.745} & {\scriptsize{}$\arrowvert H-1\rightarrow L+1\rangle$}\tabularnewline
 &  &  & {\scriptsize{}$\arrowvert H-1\rightarrow H\rangle$} &  &  & {\scriptsize{}$\arrowvert H-1\rightarrow L;H\rightarrow L+2\rangle$}\tabularnewline
\hline 
{\scriptsize{}6-311++G(3df,3pd)} & {\scriptsize{}$\parallel$} & {\scriptsize{}0.119} & {\scriptsize{}$\arrowvert H\rightarrow L\rangle$} & {\scriptsize{}$\perp$} & {\scriptsize{}0.749} & {\scriptsize{}$\arrowvert H-1\rightarrow L+1\rangle$}\tabularnewline
 &  &  & {\scriptsize{}$\arrowvert H-1\rightarrow H\rangle$} &  &  & {\scriptsize{}$\arrowvert H-1\rightarrow L;H\rightarrow L+2\rangle$}\tabularnewline
\hline 
{\scriptsize{}cc-PVDZ} & {\scriptsize{}$\parallel$} & {\scriptsize{}0.114} & {\scriptsize{}$\arrowvert H\rightarrow L+2\rangle$} & {\scriptsize{}$\perp$} & {\scriptsize{}0.736} & {\scriptsize{}$\arrowvert H-1\rightarrow L\rangle$}\tabularnewline
 &  &  & {\scriptsize{}$\arrowvert H\rightarrow L;H-1\rightarrow L\rangle$} &  &  & {\scriptsize{}$\arrowvert H-1\rightarrow L+2;H\rightarrow L+3\rangle$}\tabularnewline
\hline 
{\scriptsize{}cc-pVTZ} & {\scriptsize{}$\parallel$} & {\scriptsize{}0.120} & {\scriptsize{}$\arrowvert H\rightarrow L+2\rangle$} & {\scriptsize{}$\perp$} & {\scriptsize{}0.749} & {\scriptsize{}$\arrowvert H-1\rightarrow L\rangle$}\tabularnewline
 &  &  & {\scriptsize{}$\arrowvert H\rightarrow L;H-1\rightarrow L\rangle$} &  &  & {\scriptsize{}$\arrowvert H-1\rightarrow L+2;H\rightarrow L+3\rangle$}\tabularnewline
\hline 
{\scriptsize{}aug-cc-pVDZ} & {\scriptsize{}$\parallel$} & {\scriptsize{}0.114} & {\scriptsize{}$\arrowvert H\rightarrow L+2\rangle$} & {\scriptsize{}$\perp$} & {\scriptsize{}0.768} & {\scriptsize{}$\arrowvert H-1\rightarrow L\rangle$}\tabularnewline
 &  &  & {\scriptsize{}$\arrowvert H\rightarrow L;H-1\rightarrow L\rangle$} &  &  & {\scriptsize{}$\arrowvert H-1\rightarrow L+2;H\rightarrow L+3\rangle$}\tabularnewline
\hline 
{\scriptsize{}aug-cc-pVTZ} & {\scriptsize{}$\parallel$} & {\scriptsize{}0.120} & {\scriptsize{}$\arrowvert H\rightarrow L+2\rangle$} & {\scriptsize{}$\perp$} & {\scriptsize{}0.757} & {\scriptsize{}$\arrowvert H-1\rightarrow L\rangle$}\tabularnewline
 &  &  & {\scriptsize{}$\arrowvert H\rightarrow L;H-1\rightarrow L\rangle$} &  &  & {\scriptsize{}$\arrowvert H-1\rightarrow L+2;H\rightarrow L+3\rangle$}\tabularnewline
\hline 
\end{tabular}
\end{table}

\begin{table}[H]
\caption{Comparison of oscillator strengths and the dominant configurations
contributing to the many-particle wave functions for peaks I and the
maximum intensity peak (peak VII) of Be$_{3}^{+}$ cluster calculated
using different basis sets. The rest of the information is same as
in the caption of Table \ref{tab:osc-str-li3-linear}\label{tab:osc-strength-be3+}}

\centering{}%
\begin{tabular}{|c|c|c|c|c|c|c|}
\hline 
{\scriptsize{}Basis Set} & \multicolumn{1}{c}{} & \multicolumn{1}{c}{{\scriptsize{}Peak I}} &  & \multicolumn{1}{c}{} & \multicolumn{1}{c}{{\scriptsize{}Peak VII}} & \tabularnewline
\cline{2-7} \cline{3-7} \cline{4-7} \cline{5-7} \cline{6-7} \cline{7-7} 
 & {\scriptsize{}Polarization} & {\scriptsize{}$f$} & {\scriptsize{}Wave-function} & {\scriptsize{}Polarization} & {\scriptsize{}$f$} & {\scriptsize{}Wave-function}\tabularnewline
\hline 
\hline 
{\scriptsize{}6-311++G(2d,2p)} & {\scriptsize{}$\parallel$} & {\scriptsize{}0.172} & {\scriptsize{}$\arrowvert HF\rangle$} & {\scriptsize{}$\perp$} & {\scriptsize{}0.655} & {\scriptsize{}$\arrowvert H-2\rightarrow L+1\rangle$}\tabularnewline
 &  &  & {\scriptsize{}$\arrowvert H\rightarrow L\rangle$} &  &  & {\scriptsize{}$\arrowvert H-1\rightarrow L+2\rangle$}\tabularnewline
\hline 
{\scriptsize{}6-311++G(3df,3pd)} & {\scriptsize{}$\parallel$} & {\scriptsize{}0.184} & {\scriptsize{}$\arrowvert H\rightarrow L+1\rangle$} & {\scriptsize{}$\perp$} & {\scriptsize{}0.647} & {\scriptsize{}$\arrowvert H-2\rightarrow L;H\rightarrow L+2\rangle$}\tabularnewline
 &  &  & {\scriptsize{}$\arrowvert H-1\rightarrow L;H\rightarrow L\rangle$} &  &  & {\scriptsize{}$\arrowvert H-1\rightarrow L;H\rightarrow L+4\rangle$}\tabularnewline
\hline 
{\scriptsize{}cc-PVDZ} & {\scriptsize{}$\parallel$} & {\scriptsize{}0.178} & {\scriptsize{}$\arrowvert H\rightarrow L\rangle$} & {\scriptsize{}$\perp$} & {\scriptsize{}0.640} & {\scriptsize{}$\arrowvert H-2\rightarrow L+1\rangle$}\tabularnewline
 &  &  & {\scriptsize{}$\arrowvert H-1\rightarrow H\rangle$} &  &  & {\scriptsize{}$\arrowvert H-1\rightarrow L+2\rangle$}\tabularnewline
\hline 
{\scriptsize{}cc-pVTZ} & {\scriptsize{}$\parallel$} & {\scriptsize{}0.169} & {\scriptsize{}$\arrowvert HF\rangle$} & {\scriptsize{}$\perp$} & {\scriptsize{}0.635} & {\scriptsize{}$\arrowvert H-2\rightarrow L;H\rightarrow L+1\rangle$}\tabularnewline
 &  &  & {\scriptsize{}$\arrowvert H-1\rightarrow L;H\rightarrow L\rangle$} &  &  & {\scriptsize{}$\arrowvert H-1\rightarrow L;H\rightarrow L+2\rangle$}\tabularnewline
\hline 
{\scriptsize{}aug-cc-pVDZ} & {\scriptsize{}$\parallel$} & {\scriptsize{}0.180} & {\scriptsize{}$\arrowvert H\rightarrow L+1\rangle$} & {\scriptsize{}$\perp$} & {\scriptsize{}0.646} & {\scriptsize{}$\arrowvert H-2\rightarrow L;H\rightarrow L+2\rangle$}\tabularnewline
 &  &  & {\scriptsize{}$\arrowvert H-1\rightarrow L;H\rightarrow L\rangle$} &  &  & {\scriptsize{}$\arrowvert H-1\rightarrow L;H\rightarrow L+4\rangle$}\tabularnewline
\hline 
{\scriptsize{}aug-cc-pVTZ} & {\scriptsize{}$\parallel$} & {\scriptsize{}0.179} & {\scriptsize{}$\arrowvert H\rightarrow L+1\rangle$} & {\scriptsize{}$\perp$} & {\scriptsize{}0.657} & {\scriptsize{}$\arrowvert H-2\rightarrow L;H\rightarrow L+2\rangle$}\tabularnewline
 &  &  & {\scriptsize{}$\arrowvert H-1\rightarrow L;H\rightarrow L\rangle$} &  &  & {\scriptsize{}$\arrowvert H-1\rightarrow L;H\rightarrow L+4\rangle$}\tabularnewline
\hline 
\end{tabular}
\end{table}

\begin{table}[H]
\caption{Comparison of oscillator strengths and the dominant configurations
contributing to the many-particle wave functions for peaks I and the
maximum intensity peak (peak IV) of B$_{2}^{+}$ cluster calculated
using different basis sets. The rest of the information is same as
in the caption of Table \ref{tab:osc-str-li3-linear}\label{tab:Osc-strength-b2+}}

\centering{}%
\begin{tabular}{|c|c|c|c|c|c|c|}
\hline 
{\scriptsize{}Basis Set} & \multicolumn{1}{c}{} & \multicolumn{1}{c}{{\scriptsize{}Peak I}} &  & \multicolumn{1}{c}{} & \multicolumn{1}{c}{{\scriptsize{}Peak IV}} & \tabularnewline
\cline{2-7} \cline{3-7} \cline{4-7} \cline{5-7} \cline{6-7} \cline{7-7} 
 & {\scriptsize{}Polarization} & {\scriptsize{}$f$} & {\scriptsize{}Wave-function} & {\scriptsize{}Polarization} & {\scriptsize{}$f$} & {\scriptsize{}Wave-function}\tabularnewline
\hline 
\hline 
{\scriptsize{}6-311++G(2d,2p)} & {\scriptsize{}$\parallel$} & {\scriptsize{}0.026} & {\scriptsize{}$\arrowvert H-1\rightarrow L;H\rightarrow L\rangle$} & {\scriptsize{}$\parallel$} & {\scriptsize{}0.910} & {\scriptsize{}$\arrowvert H\rightarrow L+5\rangle$}\tabularnewline
 &  &  & {\scriptsize{}$\arrowvert H\rightarrow L+11\rangle$} &  &  & {\scriptsize{}$\arrowvert H\rightarrow L+11\rangle$}\tabularnewline
\hline 
{\scriptsize{}6-311++G(3df,3pd)} & {\scriptsize{}$\parallel$} & {\scriptsize{}0.025} & {\scriptsize{}$\arrowvert H-1\rightarrow L;H\rightarrow L\rangle$} & {\scriptsize{}$\parallel$} & {\scriptsize{}0.915} & {\scriptsize{}$\arrowvert H\rightarrow L+5\rangle$}\tabularnewline
 &  &  & {\scriptsize{}$\arrowvert H\rightarrow L+11\rangle$} &  &  & {\scriptsize{}$\arrowvert H\rightarrow L+11\rangle$}\tabularnewline
\hline 
{\scriptsize{}cc-PVDZ} & {\scriptsize{}$\parallel$} & {\scriptsize{}0.025} & {\scriptsize{}$\arrowvert H-1\rightarrow L;H\rightarrow L\rangle$} & {\scriptsize{}$\parallel$} & {\scriptsize{}0.965} & {\scriptsize{}$\arrowvert H\rightarrow L+5\rangle$}\tabularnewline
 &  &  & {\scriptsize{}$\arrowvert H\rightarrow L+5\rangle$} &  &  & {\scriptsize{}$\arrowvert H-1\rightarrow L;H\rightarrow L\rangle$}\tabularnewline
\hline 
{\scriptsize{}cc-pVTZ} & {\scriptsize{}$\parallel$} & {\scriptsize{}0.024} & {\scriptsize{}$\arrowvert H-1\rightarrow L;H\rightarrow L\rangle$} & {\scriptsize{}$\parallel$} & {\scriptsize{}0.942} & {\scriptsize{}$\arrowvert H\rightarrow L+5\rangle$}\tabularnewline
 &  &  & {\scriptsize{}$\arrowvert H\rightarrow L+5\rangle$} &  &  & {\scriptsize{}$\arrowvert H-1\rightarrow L;H\rightarrow L\rangle$}\tabularnewline
\hline 
{\scriptsize{}aug-cc-pVDZ} & {\scriptsize{}$\parallel$} & {\scriptsize{}0.028} & {\scriptsize{}$\arrowvert H-1\rightarrow L;H\rightarrow L\rangle$} & {\scriptsize{}$\parallel$} & {\scriptsize{}0.895} & {\scriptsize{}$\arrowvert H\rightarrow L+11\rangle$}\tabularnewline
 &  &  & {\scriptsize{}$\arrowvert H\rightarrow L+11\rangle$} &  &  & {\scriptsize{}$\arrowvert H\rightarrow L+5\rangle$}\tabularnewline
\hline 
{\scriptsize{}aug-cc-pVTZ} & {\scriptsize{}$\parallel$} & {\scriptsize{}0.026} & {\scriptsize{}$\arrowvert H-1\rightarrow L;H\rightarrow L\rangle$} & {\scriptsize{}$\parallel$} & {\scriptsize{}0.918} & {\scriptsize{}$\arrowvert H\rightarrow L+11\rangle$}\tabularnewline
 &  &  & {\scriptsize{}$\arrowvert H\rightarrow L+11\rangle$} &  &  & {\scriptsize{}$\arrowvert H\rightarrow L+5\rangle$}\tabularnewline
\hline 
\end{tabular}
\end{table}

\begin{table}[H]
\caption{Comparison of oscillator strengths and the dominant configurations
contributing to the many-particle wave functions for peaks I and the
maximum intensity peak (peak VIII) of B$_{3}^{+}$ cluster calculated
using different basis sets. The rest of the information is same as
in the caption of Table \ref{tab:osc-str-li3-linear}\label{tab:Osc-strength-b3+}}

\centering{}{\footnotesize{}}%
\begin{tabular}{|c|c|c|c|c|c|c|}
\hline 
{\scriptsize{}Basis Set} & \multicolumn{1}{c}{} & \multicolumn{1}{c}{{\scriptsize{}Peak I}} &  & \multicolumn{1}{c}{} & \multicolumn{1}{c}{{\scriptsize{}Peak VIII}} & \tabularnewline
\cline{2-7} \cline{3-7} \cline{4-7} \cline{5-7} \cline{6-7} \cline{7-7} 
 & {\scriptsize{}Polarization} & {\scriptsize{}$f$} & {\scriptsize{}Wave-function} & {\scriptsize{}Polarization} & {\scriptsize{}$f$} & {\scriptsize{}Wave-function}\tabularnewline
\hline 
\hline 
{\scriptsize{}6-311++G(2d,2p)} & {\scriptsize{}$\perp$} & {\scriptsize{}0.005} & {\scriptsize{}$\arrowvert H\rightarrow L\rangle$} & {\scriptsize{}$\parallel$} & {\scriptsize{}0.508} & {\scriptsize{}$\arrowvert H-1\rightarrow L;H\rightarrow L+3\rangle$}\tabularnewline
 &  &  & {\scriptsize{}$\arrowvert H-2\rightarrow L;H\rightarrow L+1\rangle$} &  &  & {\scriptsize{}$\arrowvert H-1\rightarrow L;H\rightarrow L+2\rangle$}\tabularnewline
\hline 
{\scriptsize{}6-311++G(3df,3pd)} & {\scriptsize{}$\perp$} & {\scriptsize{}0.005} & {\scriptsize{}$\arrowvert H\rightarrow L\rangle$} & {\scriptsize{}$\parallel$} & {\scriptsize{}0.514} & {\scriptsize{}$\arrowvert H-1\rightarrow L;H\rightarrow L+3\rangle$}\tabularnewline
 &  &  & {\scriptsize{}$\arrowvert H-1\rightarrow L;H\rightarrow L+1\rangle$} &  &  & {\scriptsize{}$\arrowvert H-1\rightarrow L;H\rightarrow L+2\rangle$}\tabularnewline
\hline 
{\scriptsize{}cc-PVDZ} & {\scriptsize{}$\perp$} & {\scriptsize{}0.007} & {\scriptsize{}$\arrowvert H\rightarrow L\rangle$} & {\scriptsize{}$\parallel$} & {\scriptsize{}0.480} & {\scriptsize{}$\arrowvert H-1\rightarrow L;H\rightarrow L+3\rangle$}\tabularnewline
 &  &  & {\scriptsize{}$\arrowvert H-1\rightarrow L;H\rightarrow L+1\rangle$} &  &  & {\scriptsize{}$\arrowvert H-1\rightarrow L;H\rightarrow L+3\rangle$}\tabularnewline
\hline 
{\scriptsize{}cc-pVTZ} & {\scriptsize{}$\perp$} & {\scriptsize{}0.005} & {\scriptsize{}$\arrowvert H\rightarrow L\rangle$} & {\scriptsize{}$\parallel$} & {\scriptsize{}0.491} & {\scriptsize{}$\arrowvert H-1\rightarrow L;H\rightarrow L+3\rangle$}\tabularnewline
 &  &  & {\scriptsize{}$\arrowvert H-1\rightarrow L;H\rightarrow L+1\rangle$} &  &  & {\scriptsize{}$\arrowvert H-1\rightarrow L;H\rightarrow L+2\rangle$}\tabularnewline
\hline 
{\scriptsize{}aug-cc-pVDZ} & {\scriptsize{}$\perp$} & {\scriptsize{}0.005} & {\scriptsize{}$\arrowvert H\rightarrow L\rangle$} & {\scriptsize{}$\parallel$} & {\scriptsize{}0.467} & {\scriptsize{}$\arrowvert H-1\rightarrow L;H\rightarrow L+3\rangle$}\tabularnewline
 &  &  & {\scriptsize{}$\arrowvert H-1\rightarrow L;H\rightarrow L+2\rangle$} &  &  & {\scriptsize{}$\arrowvert H-1\rightarrow L;H\rightarrow L+3\rangle$}\tabularnewline
\hline 
{\scriptsize{}aug-cc-pVTZ} & {\scriptsize{}$\perp$} & {\scriptsize{}0.006} & {\scriptsize{}$\arrowvert H\rightarrow L\rangle$} & {\scriptsize{}$\parallel$} & {\scriptsize{}0.519} & {\scriptsize{}$\arrowvert H-1\rightarrow L;H\rightarrow L+3\rangle$}\tabularnewline
 &  &  & {\scriptsize{}$\arrowvert H-1\rightarrow L;H\rightarrow L+2\rangle$} &  &  & {\scriptsize{}$\arrowvert H-1\rightarrow L;H\rightarrow L+3\rangle$}\tabularnewline
\hline 
\end{tabular}{\footnotesize\par}
\end{table}

\begin{table}[H]
\caption{Many-particle wave functions of excited states contributing to the
peaks in the optical absorption spectrum of Li$_{2}$ cluster for
aug-cc-pVTZ basis set. ’E’ corresponds to excitation energy (in eV)
of an excited state, $f$ denotes the oscillator strength for a particular
electric dipole transition. In the “|TDM|” column,we present the magnitudes
of the transition dipole moments (TDMs) to understand the extent of
coupling between the relevant excited state and the ground state.
$\parallel$ indicates photon polarization along the direction of
the molecule (longitudinal polarization), while $\perp$ indicates
polarization perpendicular to the molecular axis (transverse polarization).
’H’ and ’L’ stand for HOMO and LUMO orbitals. In the “Wave function”
column, each number inside the parentheses denotes the coefficient
of the corresponding configuration in the CI wave function. GS indicates
the ground states wave function of the cluster.\vspace{0.4cm}\label{tab:li2-wave-analysis}}

\begin{tabular}{|c|c|c|c|c|c|}
\hline 
Peak & E (eV) & $f$ & |TDM| & Polarization & Wave function\tabularnewline
\hline 
\hline 
GS &  &  &  &  & $\arrowvert HF\rangle(0.9520)$\tabularnewline
 &  &  &  &  & $\arrowvert H\rightarrow L+7;H\rightarrow L+15\rangle(0.0879)$\tabularnewline
\hline 
I & 1.83 & 0.454 & 3.184 & $\parallel$ & $\arrowvert H\rightarrow L\rangle(0.7523)$\tabularnewline
 &  &  &  &  & $\arrowvert H\rightarrow L+3\rangle(0.3959)$\tabularnewline
\hline 
II & 2.57 & 0.966 & 2.770 & $\perp$ & $\arrowvert H\rightarrow L+2\rangle(0.7046)$\tabularnewline
 &  &  &  &  & $\arrowvert H\rightarrow L+7\rangle(0.5914)$\tabularnewline
\hline 
III & 3.87 & 0.049 & 0.509 & $\perp$ & $\arrowvert H\rightarrow L+7\rangle(0.6291)$\tabularnewline
 &  &  &  &  & $\arrowvert H\rightarrow L+2\rangle(0.5439)$\tabularnewline
\hline 
V & 5.42 & 0.024 & 0.303 & $\perp$ & $\arrowvert H\rightarrow L+16\rangle(0.8509)$\tabularnewline
 &  &  &  &  & $\arrowvert H\rightarrow L+8;H\rightarrow L+16\rangle(0.1702)$\tabularnewline
\hline 
VI & 5.93 & 0.026 & 0.421 & $\parallel$ & $\arrowvert H\rightarrow L+12\rangle(0.2911)$\tabularnewline
 &  &  &  &  & $\arrowvert H\rightarrow L;H\rightarrow L+8\rangle(0.2523)$\tabularnewline
\hline 
\end{tabular}
\end{table}

\begin{table}[H]
\caption{Many-particle wave functions of excited states contributing to the
peaks in the optical absorption spectrum of Li$_{3}$ linear cluster
for aug-cc-pVTZ basis set. The rest of the information is same as
in the caption of Table \ref{tab:li2-wave-analysis}\vspace{0.4cm}\label{tab:li3-lin-wave-analysis}}

\begin{tabular}{|c|c|c|c|c|c|}
\hline 
Peak & E (eV) & $f$ & |TDM| & Polarization & Wave function\tabularnewline
\hline 
GS &  &  &  &  & $\arrowvert H-1\rightarrow H\rangle(0.9246)$\tabularnewline
 &  &  &  &  & $\arrowvert H-1\rightarrow L+13\rangle(0.0919)$\tabularnewline
\hline 
I & 0.72 & 0.106 & 2.453 & $\parallel$ & $\arrowvert HF\rangle(0.8814)$\tabularnewline
 &  &  &  &  & $\arrowvert H-1\rightarrow H;H-1\rightarrow L+5\rangle(0.1436)$\tabularnewline
\hline 
II & 1.27 & 0.503 & 4.023 & $\parallel$ & $\arrowvert H-1\rightarrow H;H-1\rightarrow L\rangle(0.5569)$\tabularnewline
 &  &  &  &  & $\arrowvert H-1\rightarrow H;H-1\rightarrow L+5\rangle(0.5102)$\tabularnewline
\hline 
III & 2.53 & 0.594 & 3.096 & $\perp$ & $\arrowvert H-1\rightarrow L+2\rangle(0.6262)$\tabularnewline
 &  &  &  &  & $\arrowvert H-1\rightarrow H;H-1\rightarrow L+11\rangle(0.3628)$\tabularnewline
\hline 
IV & 3.37 & 0.215 & 1.141 & $\perp$ & $\arrowvert H-1\rightarrow L+7\rangle(0.4255)$\tabularnewline
 &  &  &  &  & $\arrowvert H-1\rightarrow L+14\rangle(0.4237)$\tabularnewline
\hline 
V & 3.89 & 0.011 & 0.343 & $\parallel$ & $\arrowvert H-1\rightarrow L+8\rangle(0.4214)$\tabularnewline
 &  &  &  &  & $\arrowvert H-1\rightarrow H;H-1\rightarrow L+13\rangle(0.3520)$\tabularnewline
\hline 
\end{tabular}
\end{table}

\begin{table}[H]
\caption{Many-particle wave functions of excited states contributing to the
peaks in the optical absorption spectrum of Li$_{3}$ isosceles triangular
cluster for aug-cc-pVTZ basis set. The rest of the information is
same as in the caption of Table\ref{tab:li2-wave-analysis} \vspace{0.4cm}\label{tab:li3-triangular-wave-analysis}}

\begin{tabular}{|c|c|c|c|c|c|}
\hline 
Peak & E (eV) & $f$ & |TDM| & Polarization & Wave function\tabularnewline
\hline 
\hline 
GS &  &  &  &  & $\arrowvert HF\rangle(0.9102)$\tabularnewline
 &  &  &  &  & $\arrowvert H-1\rightarrow H\rangle(0.0933)$\tabularnewline
\hline 
I & 1.07 & 0.129 & 2.222 & $\parallel$ & $\arrowvert H\rightarrow L+13\rangle(0.4241)$\tabularnewline
 &  &  &  &  & $\arrowvert H\rightarrow L+1\rangle(0.4114)$\tabularnewline
\hline 
II & 1.42 & 0.020 & 0.767 & $\parallel$ & $\arrowvert H\rightarrow L+16\rangle(0.5005)$\tabularnewline
 &  &  &  &  & $\arrowvert H-1\rightarrow L\rangle(0.4115)$\tabularnewline
\hline 
III & 2.11 & 0.352 & 2.611 & $\parallel$ & $\arrowvert H-1\rightarrow H\rangle(0.5806)$\tabularnewline
 &  &  &  &  & $\arrowvert H-1\rightarrow L+15\rangle(0.2483)$\tabularnewline
\hline 
IV & 2.43 & 0.462 & 2.885 & $\parallel$ & $\arrowvert H-1\rightarrow L\rangle(0.5537)$\tabularnewline
 &  &  &  &  & $\arrowvert H\rightarrow L+5\rangle(0.3869)$\tabularnewline
\hline 
V & 2.65 & 0.088 & 1.163 & $\perp$ & $\arrowvert H\rightarrow L+2\rangle(0.5541)$\tabularnewline
 &  &  &  &  & $\arrowvert H\rightarrow L+12\rangle(0.3375)$\tabularnewline
\hline 
VI & 2.95 & 0.267 & 1.922 & $\perp$ & $\arrowvert H-1\rightarrow L+2\rangle(0.4766)$\tabularnewline
 &  &  &  &  & $\arrowvert H-1\rightarrow L+12\rangle(0.4269)$\tabularnewline
\hline 
VII & 3.20 & 0.117 & 1.223 & $\perp$ & $\arrowvert H-1\rightarrow L+2\rangle(0.3983)$\tabularnewline
 &  &  &  &  & $\arrowvert H\rightarrow L+12\rangle(0.3579)$\tabularnewline
\hline 
VIII & 3.77 & 0.016 & 0.417 & $\parallel$ & $\arrowvert H\rightarrow L+10\rangle(0.3841)$\tabularnewline
 &  &  &  &  & $\arrowvert H\rightarrow L+19\rangle(0.3349)$\tabularnewline
\hline 
IX & 4.11 & 0.029 & 0.533 & $\parallel$ & $\arrowvert H-1\rightarrow L+14\rangle(0.4375)$\tabularnewline
 &  &  &  &  & $\arrowvert H-1\rightarrow L\rangle(0.3910)$\tabularnewline
\hline 
X & 5.39 & 0.004 & 0.171 & $\parallel$ & $\arrowvert H\rightarrow L+35\rangle(0.2813)$\tabularnewline
 &  &  &  &  & $\arrowvert H\rightarrow L+37\rangle(0.2744)$\tabularnewline
\hline 
\end{tabular}
\end{table}

\begin{table}[H]
\caption{Many-particle wave functions of excited states contributing to the
peaks in the optical absorption spectrum of Li$_{4}$ cluster for
aug-cc-pVTZ basis set. The rest of the information is same as in the
caption of Table \ref{tab:li2-wave-analysis}\vspace{0.4cm}\label{tab:li4-wave-analysis}}

\begin{tabular}{|c|c|c|c|c|c|}
\hline 
Peak & E (eV) & $f$ & |TDM| & Polarization & Wave function\tabularnewline
\hline 
GS &  &  &  &  & $\arrowvert HF\rangle(0.8913)$\tabularnewline
 &  &  &  &  & $\arrowvert(H-1)\rightarrow L;H\rightarrow L+18\rangle(0.0791)$\tabularnewline
\hline 
I & 1.87 & 0.656 & 3.785 & $\parallel$ & $\arrowvert H\rightarrow L+1\rangle(0.5922)$\tabularnewline
 &  &  &  &  & $\arrowvert H\rightarrow L+9\rangle(0.4041)$\tabularnewline
\hline 
II & 2.65 & 0.660 & 3.188 & $\parallel$ & $\arrowvert H-1\rightarrow L\rangle(0.6193)$\tabularnewline
 &  &  &  &  & $\arrowvert H-1\rightarrow L+5\rangle(0.2910)$\tabularnewline
\hline 
III & 2.93 & 0.316 & 2.098 & $\perp$ & $\arrowvert H\rightarrow L+7\rangle(0.4577)$\tabularnewline
 &  &  &  &  & $\arrowvert H\rightarrow L+20\rangle(0.4174)$\tabularnewline
\hline 
IV & 3.43 & 0.045 & 0.736 & $\perp$ & $\arrowvert H-1\rightarrow L+3\rangle(0.3080)$\tabularnewline
 &  &  &  &  & $\arrowvert H\rightarrow L+7\rangle(0.2299)$\tabularnewline
\hline 
V & 3.66 & 0.103 & 1.070 & $\parallel$ & $\arrowvert H\rightarrow L+6\rangle(0.5653)$\tabularnewline
 &  &  &  &  & $\arrowvert H\rightarrow L+18\rangle(0.3856)$\tabularnewline
\hline 
VI & 4.22 & 0.077 & 0.862 & $\perp$ & $\arrowvert H-1\rightarrow L+3\rangle(0.2380)$\tabularnewline
 &  &  &  &  & $\arrowvert H\rightarrow L+20;H\rightarrow L+21\rangle(0.2227)$\tabularnewline
\hline 
VII & 4.61 & 0.063 & 0.746 & $\perp$ & $\arrowvert H-1\rightarrow L+3\rangle(0.4044)$\tabularnewline
 &  &  &  &  & $\arrowvert H-1\rightarrow L+12\rangle(0.4026)$\tabularnewline
\hline 
VIII & 5.30 & 0.012 & 0.300 & $\parallel$ & $\arrowvert H\rightarrow L+33\rangle(0.3681)$\tabularnewline
 &  &  &  &  & $\arrowvert H\rightarrow L;H\rightarrow L+6\rangle(0.2294)$\tabularnewline
\hline 
\end{tabular}
\end{table}

\begin{table}[H]
\caption{Many-particle wave functions of excited states contributing to the
peaks in the optical absorption spectrum of Be$_{2}^{+}$ cluster
for aug-cc-pVTZ basis set. The rest of the information is same as
in the caption of Table \ref{tab:li2-wave-analysis}\vspace{0.4cm}
\label{tab:be2+-wave-analysis}}

\begin{tabular}{|c|c|c|c|c|c|}
\hline 
Peak & E (eV) & $f$ & |TDM| & Polarization & Wave function\tabularnewline
\hline 
GS &  &  &  &  & $\arrowvert H\rightarrow L\rangle(0.9351)$\tabularnewline
 &  &  &  &  & $\arrowvert H-1\rightarrow L;H\rightarrow L+2\rangle(0.1684)$\tabularnewline
\hline 
I & 1.74 & 0.120 & 1.680 & $\parallel$ & $\arrowvert H\rightarrow L+2\rangle(0.8403)$\tabularnewline
 &  &  &  &  & $\arrowvert H-1\rightarrow L;H\rightarrow L\rangle(0.3820)$\tabularnewline
\hline 
II & 3.67 & 0.121 & 0.821 & $\perp$ & $\arrowvert H\rightarrow L+3\rangle(0.6705)$\tabularnewline
 &  &  &  &  & $\arrowvert H\rightarrow L+4\rangle(0.6375)$\tabularnewline
\hline 
III & 4.19 & 0.386 & 1.940 & $\parallel$ & $\arrowvert H-1\rightarrow L;H\rightarrow L\rangle(0.7268)$\tabularnewline
 &  &  &  &  & $\arrowvert H\rightarrow L+2\rangle(0.3897)$\tabularnewline
\hline 
IV & 6.01 & 0.201 & 0.827 & $\perp$ & $\arrowvert H-1\rightarrow L\rangle(0.6418)$\tabularnewline
 &  &  &  &  & $\arrowvert H-1\rightarrow L;H\rightarrow L+1\rangle(0.6116)$\tabularnewline
\hline 
V & 6.30 & 0.757 & 1.566 & $\perp$ & $\arrowvert H-1\rightarrow L\rangle(0.8762)$\tabularnewline
 &  &  &  &  & $\arrowvert H-1\rightarrow L+2;H\rightarrow L+3\rangle(0.8573)$\tabularnewline
\hline 
\end{tabular}
\end{table}

\begin{table}[H]
\caption{Many-particle wave functions of excited states contributing to the
peaks in the optical absorption spectrum of Be$_{3}^{+}$ cluster
for aug-cc-pVTZ basis set. The rest of the information is same as
in the caption of Table \ref{tab:li2-wave-analysis}\vspace{0.4cm}
\label{tab:be3+-wave-analysis}}

\begin{tabular}{|c|c|c|c|c|c|}
\hline 
Peak & E (eV) & $f$ & |TDM| & Polarization & Wave function\tabularnewline
\hline 
GS &  &  &  &  & $\arrowvert H\rightarrow L\rangle(.8776)$\tabularnewline
 &  &  &  &  & $\arrowvert H-1\rightarrow L+1;H\rightarrow L\rangle(.1913)$\tabularnewline
\hline 
I & 1.02 & 0.179 & 2.678702 & $\parallel$ & $\arrowvert H\rightarrow L+1\rangle(0.8200)$\tabularnewline
 &  &  &  &  & $\arrowvert H-1\rightarrow L;H\rightarrow L\rangle(0.3257)$\tabularnewline
\hline 
II & 3.16 & 0.433 & 2.365586 & $\parallel$ & $\arrowvert H-1\rightarrow L;H\rightarrow L\rangle(0.6152)$\tabularnewline
 &  &  &  &  & $\arrowvert H-1\rightarrow L+1;H\rightarrow L+1\rangle(0.4714)$\tabularnewline
\hline 
III & 3.66 & 0.028 & 0.563211 & $\perp$ & $\arrowvert H-2\rightarrow L;H\rightarrow L+2\rangle(0.5209)$\tabularnewline
 &  &  &  &  & $\arrowvert H\rightarrow L+17\rangle(0.4125)$\tabularnewline
\hline 
IV & 4.87 & 0.020 & 0.407823 & $\perp$ & $\arrowvert H-1\rightarrow L+1;H\rightarrow L+2\rangle(0.5211)$\tabularnewline
 &  &  &  &  & $\arrowvert H\rightarrow L+17\rangle(0.3217)$\tabularnewline
\hline 
V & 5.40 & 0.204 & 1.242734 & $\parallel$ & $\arrowvert H-2\rightarrow L;H\rightarrow L+1\rangle(0.6337)$\tabularnewline
 &  &  &  &  & $\arrowvert H-1\rightarrow L;H\rightarrow L\rangle(0.3104)$\tabularnewline
\hline 
VI & 5.85 & 0.054 & 0.612656 & $\perp$ & $\arrowvert H-1\rightarrow L;H\rightarrow L+4\rangle(0.4315)$\tabularnewline
 &  &  &  &  & $\arrowvert H-1\rightarrow L;H-1\rightarrow L+2;H\rightarrow L\rangle(0.2864)$\tabularnewline
\hline 
VII & 6.52 & 0.657 & 2.028291 & $\perp$ & $\arrowvert H-2\rightarrow L;H\rightarrow L+2\rangle(0.4960)$\tabularnewline
 &  &  &  &  & $\arrowvert H-1\rightarrow L;H\rightarrow L+4\rangle(0.4653)$\tabularnewline
\hline 
\end{tabular}
\end{table}

\begin{table}[H]
\caption{Many-particle wave functions of excited states contributing to the
peaks in the optical absorption spectrum of B$_{2}^{+}$ cluster for
aug-cc-pVTZ basis set. The rest of the information is same as in the
caption of Table \ref{tab:li2-wave-analysis}\vspace{0.4cm} \label{tab:b2+-wave-analysis}}

\begin{tabular}{|c|c|c|c|c|c|}
\hline 
Peak & E (eV) & $f$ & |TDM| & Polarization & Wave function\tabularnewline
\hline 
GS &  &  &  &  & $\arrowvert H\rightarrow L\rangle(.9033)$\tabularnewline
 &  &  &  &  & $\arrowvert H-2\rightarrow L;H-1\rightarrow L+2\rangle(.1271)$\tabularnewline
\hline 
I & 3.66 & 0.026 & 0.537 & $\parallel$ & $\arrowvert H-1\rightarrow L;H\rightarrow L\rangle(0.7250)$\tabularnewline
 &  &  &  &  & $\arrowvert H\rightarrow L+11\rangle(0.2622)$\tabularnewline
\hline 
II & 4.80 & 0.029 & 0.496 & $\parallel$ & $\arrowvert H-1\rightarrow L+1;H\rightarrow L+1\rangle(0.5092)$\tabularnewline
 &  &  &  &  & $\arrowvert H-1\rightarrow H\rangle(0.4944)$\tabularnewline
\hline 
III & 5.99 & 0.019 & 0.364 & $\perp$ & $\arrowvert H-1\rightarrow L;H\rightarrow L+2\rangle(0.6138)$\tabularnewline
 &  &  &  &  & $\arrowvert H-2\rightarrow L\rangle(0.4445)$\tabularnewline
\hline 
IV & 7.04 & 0.918 & 2.307 & $\parallel$ & $\arrowvert H\rightarrow L+11\rangle(0.4808)$\tabularnewline
 &  &  &  &  & $\arrowvert H\rightarrow L+5\rangle(0.4713)$\tabularnewline
\hline 
V & 7.65 & 0.009 & 0.216 & $\perp$ & $\arrowvert H-2\rightarrow L\rangle(0.5322)$\tabularnewline
 &  &  &  &  & $\arrowvert H-1\rightarrow L;H\rightarrow L+2\rangle(0.4844)$\tabularnewline
\hline 
\end{tabular}
\end{table}

\begin{table}[H]
\caption{Many-particle wave functions of excited states contributing to the
peaks in the optical absorption spectrum of B$_{3}^{+}$ cluster for
aug-cc-pVTZ basis set. The rest of the information is same as in the
caption of Table \ref{tab:li2-wave-analysis}\vspace{0.4cm} \label{tab:b3+-wave-analysis}}

\begin{tabular}{|c|c|c|c|c|c|}
\hline 
Peak & E (eV) & $f$ & |TDM| & Polarization & Wave function\tabularnewline
\hline 
GS &  &  &  &  & $\arrowvert HF\rangle(0.8535)$\tabularnewline
 &  &  &  &  & $\arrowvert H-1\rightarrow L+1\rangle(0.1393)$\tabularnewline
\hline 
I & 0.84 & 0.006 & 0.524 & $\perp$ & $\arrowvert H\rightarrow L\rangle(0.8572)$\tabularnewline
 &  &  &  &  & $\arrowvert H-1\rightarrow L;H\rightarrow L+2\rangle(0.1319)$\tabularnewline
\hline 
II & 3.22 & 0.083 & 0.726 & $\parallel$ & $\arrowvert H-1\rightarrow L\rangle(0.8246)$\tabularnewline
 &  &  &  &  & $\arrowvert H-1\rightarrow L;H-1\rightarrow L+1\rangle(0.1642)$\tabularnewline
\hline 
III & 5.01 & 0.053 & 0.463 & $\parallel$ & $\arrowvert H-1\rightarrow L;H-1\rightarrow L\rangle(0.5482)$\tabularnewline
 &  &  &  &  & $\arrowvert H\rightarrow L;H\rightarrow L+1\rangle(0.3184)$\tabularnewline
\hline 
IV & 5.45 & 0.013 & 0.217 & $\parallel$ & $\arrowvert H\rightarrow L;H\rightarrow L+1\rangle(0.5721)$\tabularnewline
 &  &  &  &  & $\arrowvert H\rightarrow L;H\rightarrow L+2\rangle(0.5675)$\tabularnewline
\hline 
V & 6.03 & 0.026 & 0.295 & $\parallel$ & $\arrowvert H\rightarrow L+3\rangle(0.3937)$\tabularnewline
 &  &  &  &  & $\arrowvert H-1\rightarrow L+2\rangle(0.3723)$\tabularnewline
\hline 
VI & 7.16 & 0.026 & 0.382 & $\parallel$ & $\arrowvert H\rightarrow L+1;H\rightarrow L+2\rangle(0.5866)$\tabularnewline
 &  &  &  &  & $\arrowvert H\rightarrow L;H\rightarrow L+1\rangle(0.4308)$\tabularnewline
\hline 
VII & 8.08 & 0.059 & 0.380 & $\parallel$ & $\arrowvert H-1\rightarrow L;H-1\rightarrow L+1\rangle(0.4529)$\tabularnewline
 &  &  &  &  & $\arrowvert H-1\rightarrow L;H-1\rightarrow L+2\rangle(0.3162)$\tabularnewline
\hline 
VIII & 8.85 & 0.519 & 1.091 & $\parallel$ & $\arrowvert H\rightarrow L+3;H-1\rightarrow L\rangle(0.3762)$\tabularnewline
 &  &  &  &  & $\arrowvert H\rightarrow L+3;H-1\rightarrow L\rangle(0.3748)$\tabularnewline
\hline 
\end{tabular}
\end{table}

\begin{figure}[H]
\includegraphics[scale=0.3]{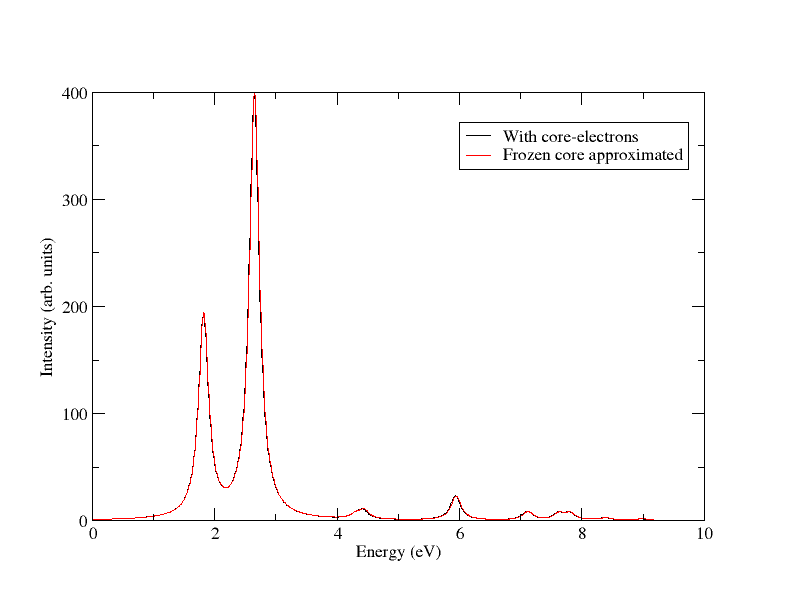}

\caption{Validation of frozen core approximation for the optical absorption
spectra of Li$_{2}$ cluster employing QCI method. }
\end{figure}

\begin{figure}[H]
\includegraphics[scale=0.3]{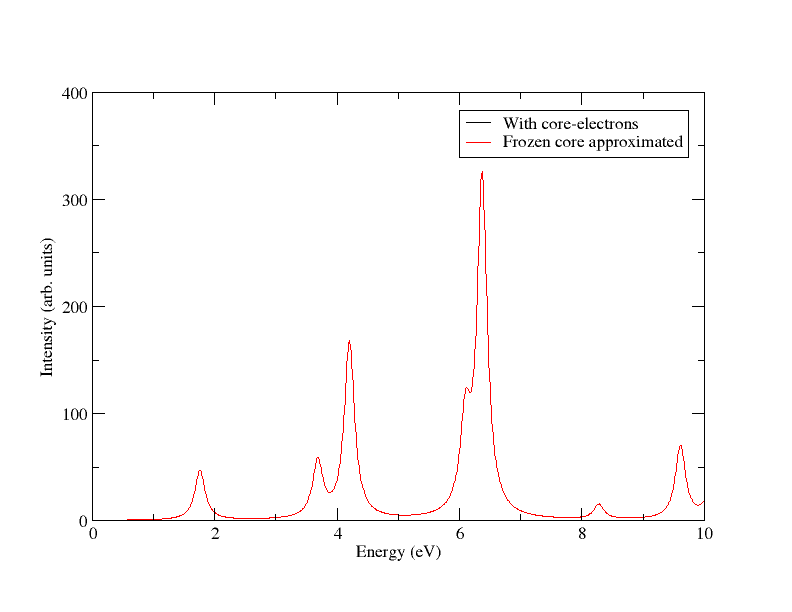}

\caption{Validation of frozen core approximation for the optical absorption
spectra of Be$_{2}^{+}$ cluster employing QCI method.}
\end{figure}

\begin{figure}[H]
\includegraphics[scale=0.3]{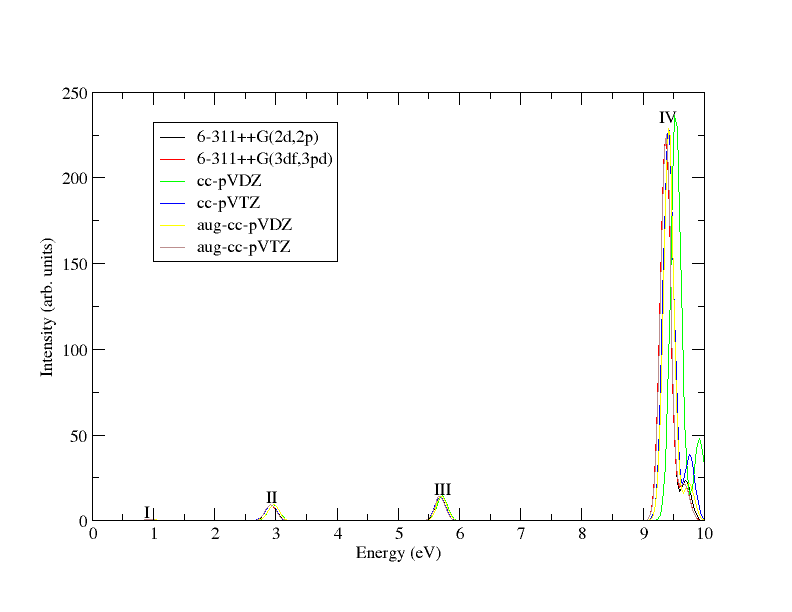}

\caption{Optical absorption spectra of B$_{3}^{+}$ cluster computed using
various basis sets employing B3LYP functional and TD-DFT method. }
\end{figure}